\documentclass[%
 aip,
rsi,%
 amsmath,amssymb,
 reprint,%
]{revtex4-1}

\usepackage{graphicx}
\usepackage{dcolumn}
\usepackage{bm}
\usepackage[usenames, dvipsnames]{color}
\usepackage[colorinlistoftodos]{todonotes}
\usepackage[capitalise]{cleveref}
\usepackage{apptools}
\usepackage{appendix}
\crefname{appsec}{Appendix}{Appendices}
\crefname{subappendix}{Appendix}{Appendices}
\usepackage{SIunits}

\newcommand{\rnorm}{(\rho_{0})}
\newcommand{\sample}{BaFe$_{1.975}$Co$_{0.025}$As$_2$ }
\newcommand{\sampleNS}{BaFe$_{1.975}$Co$_{0.025}$As$_2$}
\begin{document}

\preprint{}

\title[Measurement of Elastoresistivity at Finite Frequency by Amplitude Demodulation]{Measurement of Elastoresistivity at Finite Frequency by Amplitude Demodulation}

\author{Alexander T. Hristov}
\affiliation{Geballe Laboratory for Advanced Materials, Stanford University, Stanford, CA 94305}
\affiliation{Department of Physics, Stanford University, Stanford, CA 94305}
\author{Johanna C. Palmstrom}
\author{Joshua A. W. Straquadine}
\author{Tyler A. Merz}
\author{Harold Y. Hwang}
\author{Ian R. Fisher}
\affiliation{Geballe Laboratory for Advanced Materials, Stanford University, Stanford, CA 94305}
\affiliation{Department of Applied Physics, Stanford University, Stanford, CA 94305}

\date{\today}

\begin{abstract}
Elastoresistivity, the relation between resistivity and strain, can elucidate subtle properties of the electronic structure of a material and is an increasingly important tool for the study of strongly correlated materials.
To date, elastoresistivity measurements have been predominantly performed with quasi-static (DC) strain.
In this work, we demonstrate a method for using AC strain in elastoresistivity measurements.
A sample experiencing AC strain has a time-dependent resistivity, which modulates the voltage produced by an AC current; this effect produces time-dependent variations in resisitivity that are directly proportional to the elastoresistivity, and which can be measured more quickly, with less strain on the sample, and with less stringent requirements for temperature stability than the previous DC technique.
Example measurements between 10~Hz and 3~kHz are performed on a material with a large, well-characterized and temperature dependent elastoresistivity: the representative iron-based superconductor \sampleNS. 
These measurements yield a frequency independent elastoresistivity and reproduce results from previous DC elastoresistivity methods to within experimental accuracy. 
We emphasize that the dynamic (AC) elastoresistivity is a distinct material-specific property that has not previously been considered.
\end{abstract}

\maketitle

\section{Introduction}

Measuring strain-induced changes in electronic properties can reveal details of a material's underlying electronic structure.
Many such techniques were originally developed for the study of semiconductors,\cite{Sun_Strain_2010} but they have recently been adopted and improved in the study of correlated electron materials.
This advancement in experimental techniques has been motivated by numerous discoveries of electronic states that break rotational symmetry.\cite{ Borzi_2007, Chu_InPlane_2010, Lester_2015,  Feldman_Observation_2016, Ronning_Electronic_2017, Wu_Spontaneous_2017}
Resistivity measurements are very sensitive to electronic anisotropies of a Fermi surface and strain which breaks appropriate symmetries can be used as a conjugate field for an electronic order parameter.\cite{Shapiro_Symmetry_2015}
For these reasons, measurements of the elastoresistivity have been used to address open questions about the fluctuations of electronic nematic order\cite{Chu_Divergent_2012, Kuo_Ubiquitous_2016, Watson_Emergence_2015, Hosoi_Nematic_2016, Kuo_Measurement_2013,Tanatar_Origin_2016} and have also been extended to identify more subtle forms of compound order which break additional symmetries.\cite{Riggs_2015}

Continuous improvement to the technique of elastoresistivity opens new avenues for experimental investigation.
For example, recent technical advances enable greater strain at low temperatures while reducing unwanted strains from thermal expansion.\cite{Hicks_Piezoelectric_2014}
Other developments have enabled accurate decomposition of the elastoresistive response into distinct symmetry channels,\cite{Kuo_Measurement_2013,Shapiro_Symmetry_2015, Shapiro_Measurement_2016} enabling detection of more subtle strain-induced resistivity changes.\cite{Palmstrom_Critical_2017}
The present work continuously spans the frequency regime from the quasistatic limit to higher frequencies, providing access to the dynamical elastoresistivity as well as a method for improving the signal to noise ratio in the quasistatic limit.

DC elastoresistivity techniques operate by measuring the resistivity in each quasistatic strain environment between successive step changes in strain. 
DC elastoresistivity information is obtained through relevant fit parameters from a regression, and the process is repeated for a set of temperatures over the span of a few days.\cite{Chu_Divergent_2012}
In contrast, the AC technique presented in this paper directly produces a sinusoidal voltage with amplitude proportional to the elastoresistivity of the sample; this signal occurs at sideband frequencies equal to the sum and difference of the AC strain frequency and AC current frequency.
By exploiting the power of lock-in techniques to measure these sideband voltages, a measurement may be performed much more quickly, which has a number of important consequences.
First, a full temperature dependence of the elastoresistivity at a single frequency can be acquired while continuously varying the cryostat temperature between 4~K and 300~K in the space of a few hours. 
This is comparable to the time required for a simple resistivity measurement, and significantly less than the several days necessary for a DC elastoresistivity technique to produce the same quality of data over the same temperature range.
Second, since each data point is acquired in a very short time, the technique has much less stringent requirements for temperature stability than DC techniques and can provide improved signal to noise ratios for all materials; this is especially critical for materials with small elastoresistivity coefficients and/or strongly-temperature dependent resistivities.
Third, the technique can operate with considerably less strain on the sample, which can be of practical importance for materials with low mechanical yield points.
Fourth, the reduced measurement duration allows the use of strain in tandem with other experimental apparatuses that operate on short time scales, such as pulsed magnets.
Most significantly, though, this technique also enables investigations of the elastoresistivity response to dynamical strain, which is a distinct material property that has not been previously investigated.

\begin{figure*}[!ht]
\begin{center}
\includegraphics[width=\linewidth]{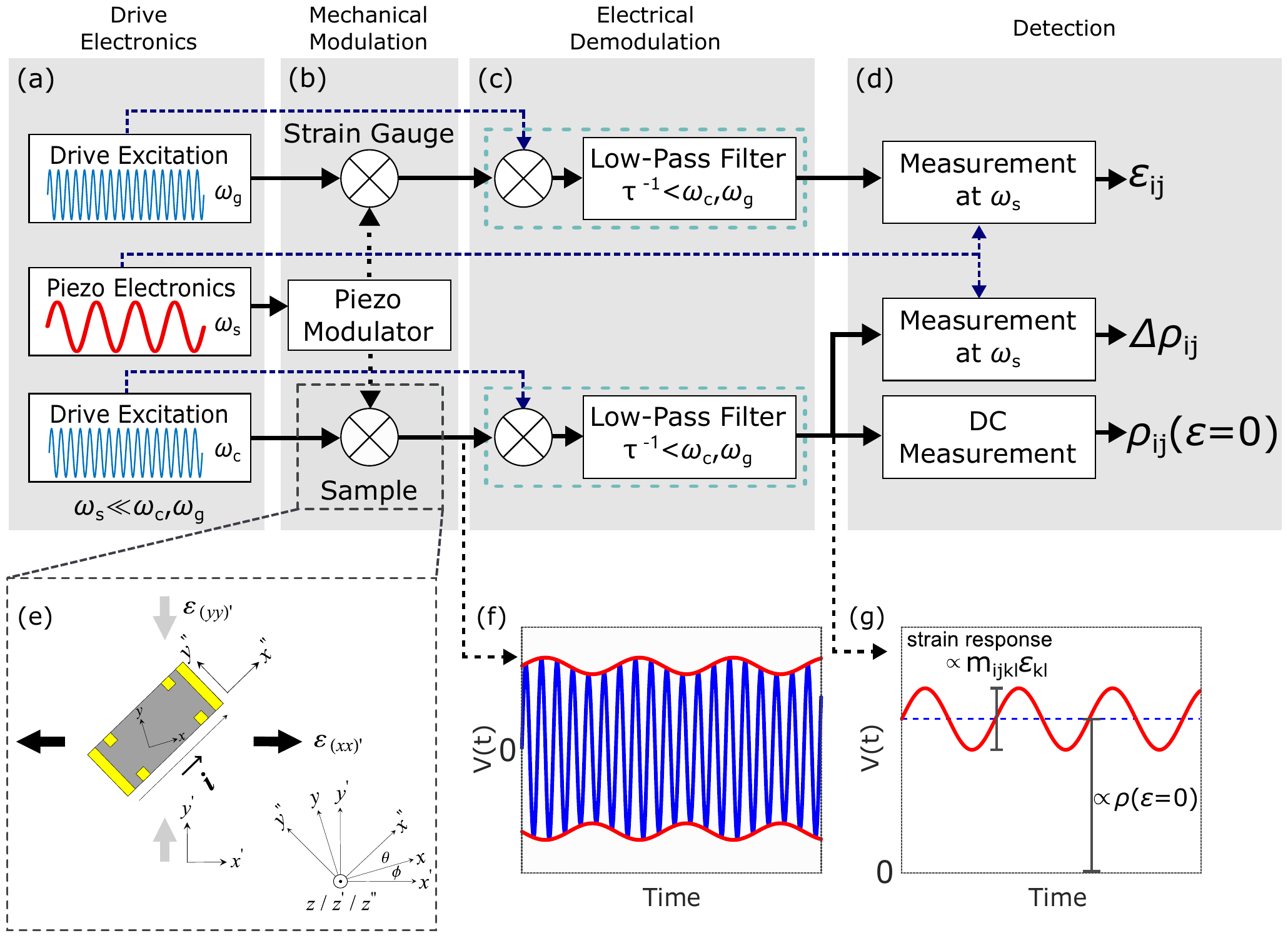}
\caption{\small \sl 
Schematic diagram illustrating the basic components and principles of the amplitude demodulation technique for measuring elastoresistivity.
Both sample and strain gauge are (a) electronically excited with sinusoidal currents and  (b) mechanically modulated by a piezoelectric device.
The oscillating strain experienced by both the sample and strain gauge will produce in each a time-varying resistance that modulates the amplitude of the voltage created by an excitation current.
An electronic mixer circuit (c) then demodulates and filters the signal (with filter time constant $\tau$ set to remove higher harmonics of the current frequency), moving these sidebands to $\omega_s$ for detection (d).  
Blue dashed lines represent reference frequencies used in demodulation measurements.
(e) Schematic of a sample (dark rectangle) prepared for an elastoresistance measurement, showing contacts (light rectangles) for four-point resistance measurements. 
The three relevant coordinate frames for these measurements are those of the primitive crystal unit cell (unprimed axes), the normal strain frame (i.e. there is zero shear with respect to the primed axes), and the current direction (double primed axes). 
The particular orientations of strain, crystal axes, applied current, and voltage contact placement must be carefully chosen in order to isolate the desired elastoresistivity component $m_{ijkl}$, but the principles of the amplitude demodulation technique presented here can be applied to any configuration.
(f) The voltage expected from a sample or resistive strain gauge modulated by strain shows fast oscillations at frequency $\omega_c$ or $\omega_g$ respectively, with a slowly varying envelope due to strain, shown in red, at frequency $\omega_s$.
(g) The voltage signal following demodulation consists of a DC component proportional to the resistivity of the sample and a time-varying component at $\omega_s$, which is the proportional to the elastoresistivity.
\label{fig:schematic}  
}
\end{center}  
\end{figure*} 

In the present paper, we focus exclusively on the linear elastoresistivity response of a material: an induced strain varying with angular frequency $\omega$ gives rise to a change in the sample resistivity at the same frequency.
The linear elastoresistivity tensor, $m_{ijkl}(\omega)$, quantifies this relation between strain and a normalized change in resistivity, in the limit of vanishing strain: 
\begin{equation}
\left( \frac{\Delta \rho}{\rho_0}\right)_{ij} (\omega)=\frac{\rho_{ij} (\omega)- \rho_{ij}(\varepsilon=0)}{ \rnorm_{ij}} = \sum_{kl} m_{ijkl}(\omega)\varepsilon_{kl}(\omega),
\label{eq:ERdefinition}
\end{equation}
where $\rnorm_{ij}$ is a normalization constant that depends on the crystal structure and which can be obtained from resistivity measurements on an unstrained sample,
\footnote{
For a cubic material, the normalization constant is no more than the isotropic unstrained resistivity: $\rnorm_{ij}=\rho_{ii}(\varepsilon=0)$.
However, this definition has an important subtlety for off diagonal terms.
Though often neglected, $\rho_{i\neq j}$ terms do arise in the absence of a magnetic field due to shear strains, and it is also appropriate to normalize them by the isotropic resistivity, thus the distinction between $\rnorm_{ij}$ and $\rho_{ij}(\varepsilon=0)$. 
For a tetragonal material, this idea extends to the in-plane elements discussed in \cref{demo}:  $\rnorm_{xx}=\rnorm_{xy} =\rnorm_{yy}=\rho_{xx}(\varepsilon=0)$.
The out-of-plane normalization conventions, not relevant to the work here, are discussed in ref. \onlinecite{Shapiro_Symmetry_2015}
} 
and $i$, $j$, $k$, and $l$ refer to the axes of any appropriate Cartesian reference frame.
Since the frequencies of strain modulation and resistivity modulation are identical to each other in the linear response regime, the frequency labels of $m_{ijkl}(\omega)$, $\varepsilon_{kl}(\omega)$, and $\left( \Delta \rho / \rho_0\right)_{ij} (\omega)$ can be dropped with no loss of clarity.

A schematic diagram of the technique is presented in \cref{fig:schematic}.
The sample experiences an oscillating strain at finite drive frequency $\omega_s$, causing the resistivity to vary at the same frequency.
When AC current at frequency $\omega_c$  is passed through the sample, the voltage difference across the sample displays the amplitude modulation shown in \cref{fig:schematic}(f).
This modulation arises because the voltage is the product of a sinusoidal current and a sinusoidally varying resistance.
Such a product of sine waves has harmonic content at sideband frequencies $\omega_\pm=|\omega_c \pm \omega_s|$.
We demonstrate in \cref{theory} how to detect signals at these sideband frequencies and extract elastoresistivity information.

The method described here is agnostic to the relative orientation of the excitation current, and contact geometry relative to the crystalline axes, as well as the relative orientation, symmetry character, and even the physical source of the sample strain.
These choices depend on which components of the elastoresistivity tensor one wishes to measure in the sample under study.
The technique is also insensitive to how one obtains an accurate measurement of the strain on the sample.
For simplicity, we focus on the common case of a resistive strain gauge.
Other strain sensors could also used, including fiber Bragg gratings and capacitive sensors; however, the strain measurement from the sensor must have sufficient bandwidth to measure amplitude of the AC strain.
The strain sensor can be substituted for a force sensor in order to design a similar dynamic piezoresistance measurement. 

\Cref{theory} introduces the basic principles for measuring dynamical elastoresistivity using AC currents and AC strain.
\Cref{StrainApplication} presents the necessary electrical and mechanical components that are required to induce strain in a sample at finite frequency and cryogenic temperatures.
\Cref{sec:electronics} details specific implementations to detect the elastoresistance signal.
\Cref{error} characterizes the sources of error in the measurement.
\Cref{demo} presents an example measurement of the elastoresistivity component $m_{xyxy}$ in a material with a large electronic nematic susceptibility, BaFe$_{1.975}$Co$_{0.025}$As$_2$, for which the DC elastoresistivity has previously been measured.\cite{Kuo_Measurement_2013}

\section{Basic Principles of Elastoresistivity Measurements Using AC Strain and AC Current}
\label{theory}

From the definition of elastoresistivity in \cref{eq:ERdefinition}, it can be shown that a sample experiencing a time-varying strain ${\varepsilon_{kl}(t)=\varepsilon^0_{kl}\sin\left(\omega_{s} t \right)}$ will exhibit a time-varying resistivity:
\begin{equation}
\rho_{ij}(t)=\rho_{ij}(\varepsilon=0) + \rnorm_{ij}\sum_{kl} m_{ijkl}  \varepsilon^0_{kl}\sin\left(\omega_{s} t +\phi_{kl} \right).
\label{eq:resistivity}
\end{equation}
At finite frequency, a phase-shift between the resistivity change and the applied strain would be captured by the phase $\phi_{kl}$.
For the specific materials considered in \cref{demo}, the elastoresistance is wholly real (i.e. ${\phi_{kl}=0}$) for frequencies up to 3~kHz. 
We neglect this phase shift to simplify subsequent discussion, though this will not be appropriate for all materials.
All subsequent expressions can be generalized to include this phase shift where necessary.


A current at frequency $\omega_{c}$ and amplitude $I_0$ can be passed through a sample with time varying resistivity specified in \cref{eq:resistivity}. 
Without loss of generality, the current can be taken along the crystallographic axes of the sample (see \cref{fig:schematic} (e)), so that the time-varying voltage, $V^{sample}_{ij}$, can be expressed as,
\begin{multline}
V^{sample}_{ij}(t) = I_0 F_{ij} \rho_{ij}(\varepsilon=0) \sin(\omega_c t) \\
+ \frac{1}{2}I_0 \rnorm_{ij} F_{ij}\sum_{kl}m_{ijkl}\varepsilon_{kl}^0\left( \cos(\omega_{c-}t) -\cos(\omega_{c+}t) \right)\label{eq:SampleSidebands}
\end{multline}
where $\omega_{c\pm}=\omega_{c}\pm\omega_{s}$ are the sideband frequencies, $F_{ij}$ is a numerical factor that relates resistivity to resistance in a particular geometry.
This signal is schematically shown in \cref{fig:schematic} (f). 
If a simultaneous measurement is also performed to measure $\rnorm_{ij}$, then comparison of \cref{eq:ERdefinition,eq:SampleSidebands} reveals the elastoresistive response can be extracted according to 
\begin{equation} \label{eq:deltarho}
\left(\frac{\Delta\rho}{\rho_0}\right)_{ij} = \frac{2\tilde{V}_{ij}(\omega_{c\pm})}{I_0 F_{ij} \rnorm_{ij}}
\end{equation}
where $\tilde{V_{ij}}$ represents the Fourier transform of \cref{eq:SampleSidebands}.

The same modulation concept can also be applied to extract the strain from a resistive strain gauge which experiences the same strain environment as the sample (oscillating at $\omega_s$). 
Unlike a sample, which is sensitive to all strains, a strain gauge is typically designed to be sensitive to deformation along a single axis, typically aligned to one of the primary axes of the piezoelectric stack.
As this axis need not be aligned with the crystal axes, we denote this coordinate system by primed indices $i^\prime$, as shown in \cref{fig:schematic} (e).
To prevent interference between sample and strain gauge, it is best to drive the strain gauge with an AC current of a different frequency $\omega_g$, resulting in a voltage
\begin{multline}
V^{\text{SG}}_{i^\prime i^\prime}(t) = I^{SG}_0 R^{SG}_0 \sin(\omega_g t)\\ + \frac{1}{2}I^{SG}_0\frac{dR}{d\varepsilon_{i^\prime i^\prime}}\varepsilon_{i^\prime i^\prime}\left( \cos(\omega_{g-}t) -\cos(\omega_{g+}t) \right)\label{eq:SGSidebands}
\end{multline}
where ${dR}/{d\varepsilon_{i^\prime i^\prime}}$ denotes the change in resistance of the gauge due to strain, typically supplied by the gauge manufacturer, and $\omega_{g\pm}=\omega_g \pm \omega_s$. 
From this voltage, one can obtain the strain according to 
\begin{equation} \label{eq:strainsig}
\varepsilon_{i^\prime i^\prime} = \frac{2 \tilde{V}^{\text{SG}}_{i^\prime i^\prime}(\omega_{g\pm}) }{I_0^{\text{SG}} \frac{dR}{d\varepsilon_{i^\prime i^\prime}}},
\end{equation}
where $\tilde{V_{ij}^\text{SG}}$ represents the Fourier transform of \cref{eq:SGSidebands}.
From a combination of enough such measurements, and \cref{eq:deltarho,eq:strainsig} it is possible to isolate elastoresistivity coefficients, $m_{ijkl}$.

In principle, this scheme can be extended to the detection of recently investigated nonlinear elastoresistivity.\cite{Palmstrom_Critical_2017}
Nonlinear elastoresistivity terms proportional to $\varepsilon^n$ would show up as higher-order sidebands, at $|\omega_c \pm n \omega_s|$ for sufficiently large strains.
Therefore after demodulation, the nonlinear elastoresistive response can be detected by locking into the $n^{th}$ harmonic of $\omega_s$.
However, any strain apparatus will show some nonlinearity with applied voltage, so a subtraction would be necessary to remove linear elastoresistance response to higher harmonic distortions present in the strain.

\section{Experimental details}
\subsection{Inducing Strain At High Frequency and Cryogenic Temperatures}
\label{StrainApplication}

There are many ways to mechanically deform samples, but piezoelectric (PE) stacks  manufactured from lead zirconate titanate or similar materials are particularly well-suited for the present technique: they are designed for use at cryogenic temperature and allow strain to be tuned continuously \textit{in situ}.
Piezoelectric stacks can readily drive oscillatory strain changes at frequencies up to the stack's mechanical resonant frequency; for commercially available stacks this can reach beyond 30~kHz.
Samples can either be directly adhered onto the side of a piezoelectric stack,\cite{Chu_Divergent_2012} or mounted to span the space between two plates that are actuated by piezoelectric stacks.\cite{Hicks_Piezoelectric_2014}

Despite the advantages of using PE stacks, three intrinsic properties of the stacks themselves must be correctly managed to enable an accurate measurement of elastoresistivity at finite frequencies.
First, the large capacitances (of order $1$~\micro\farad) of many commercially available stacks necessitates careful consideration of the driving circuitry to prevent the resulting strain becoming either diminished or harmonically distorted at high frequency.
Secondly, inefficiencies in PE stacks can result in significant heating, even for displacements well below the voltage limit on the stack. 
Finally, the electrical and mechanical properties mentioned above depend heavily on temperature.

Due to the high voltages ($>$25~V) necessary to drive maximum displacements of PE stacks, high-voltage amplifiers are often employed in elastoresistivity experiments.
To drive a sine wave at angular frequency $\omega_s$ and peak voltage $V_{p}$ into a PE stack with capacitance $C$, the driving amplifier must be able to source a peak current of
\begin{equation} \label{eq:peakcurrent}
	I_{p} = V_{p} \omega_s C
\end{equation}
For example, when the piezoelectric stack used in this work (Piezomechanik ``PSt150/5x5/7 cryo 1'', for which $C \approx$ 800~nF at 300~K) is driven at 1~kHz by a 25~V peak-to-peak voltage, a current $I_{p}>$60~mA is needed at room temperature. 
Amplifiers designed for low frequency applications, like piezoelectric positioning stages, are not generally capable of sourcing such large currents or may otherwise attenuate their voltage output at high frequency to improve the stability of the piezoelectric displacement.
Exceeding the maximum current causes both waveform distortion and a reduction in the amplitude of applied voltage (and therefore the strain) on the piezo device.
Several devices were considered for driving the piezoelectric stack, and characterization of these is shown in \cref{appendix:amplifiers}.
The Tegam 2350 amplifier characterized in the appendix is used for all the data shown elsewhere in this manuscript, as it is best capable of driving piezoelectric devices at frequencies beyond 1~kHz.

Driving the piezoelectric stacks at high frequencies and large amplitudes can produce a large heat load on the attached sample.
The piezoelectric material used here, lead zirconate titanate, has a thermal conductivity less than 1.1~Wm$^{-1}$K$^{-1}$, which is considerably less than the metals typically used for mounting samples.\cite{Yarlagadda_Low_1995}
This has two important consequences.
First, a piezoelectric provides little thermal anchoring to the cryostat, such that samples must rely more on their wires and coupling to any exchange gas for cooling, and there is greater chance of Joule self-heating.
Second, heat generated near the sample from dissipation in an oscillating piezoelectric is transferred to the sample rather than  dissipated to another part of the cryostat because the heat does not flow easily through the piezoelectric.
Computing the exact amount of heating is complicated as it depends on multiple factors including the exchange gas, the electrical contacts to the sample and the temperature dependence of many properties of the piezoelectric.
We instead take a practical approach to measure the effect using the $\rho_{xx}$ component of the sample resistivity as a thermometer.
For the 25~V peak-to-peak excitation used in this paper, self-heating of the piezoelectric increases the sample temperature by approximately 2~K at 1.5~kHz, and increases sharply (to $>$ 15~K at some temperatures) for higher frequencies, as detailed in \cref{app:Rxx}.
Where the sample resistivity is not sufficiently temperature dependent to apply this technique, a secondary thermometer should be included on the piezoelectric stack to correct for this heating or the driving amplitude must be reduced.

Many properties of piezoelectric stacks are strongly temperature dependent, so tests for heating or amplifier performance may need to be carried out over the entire range of measurement temperatures.
For the stack chosen here, the capacitance is one tenth as large at 20~K as at 300~K, which can ameliorate some of the difficulties if driving the system with constant voltage amplitude at all frequencies and temperatures.
As stated in \cref{eq:peakcurrent} the current needed from the amplifier decreases, and since the heat load is 
\begin{equation}
Q=V^2_{p} \omega_s^2 C^2 R_{p},
\end{equation}
where $R_{p}$ is the effective series resistance of the piezoelectric device, the heating also decreases if all other elements are held constant.
However, the displacement per volt also decreases at cryogenic temperatures. 
Careful characterization is therefore necessary to optimize the measurement parameters in light of these competing effects on amplifier performance, as elaborated in \cref{app:strainpervolt,app:Capacitance,app:Poisson,app:heat,app:Rxx}.

\subsection{Electrical Demodulation and Detection Circuits for Extracting Elastoresistivity from Sideband Frequencies}
\label{sec:electronics}

\begin{figure*}[!ht]  
\begin{center}  
\includegraphics[width=0.8\linewidth]{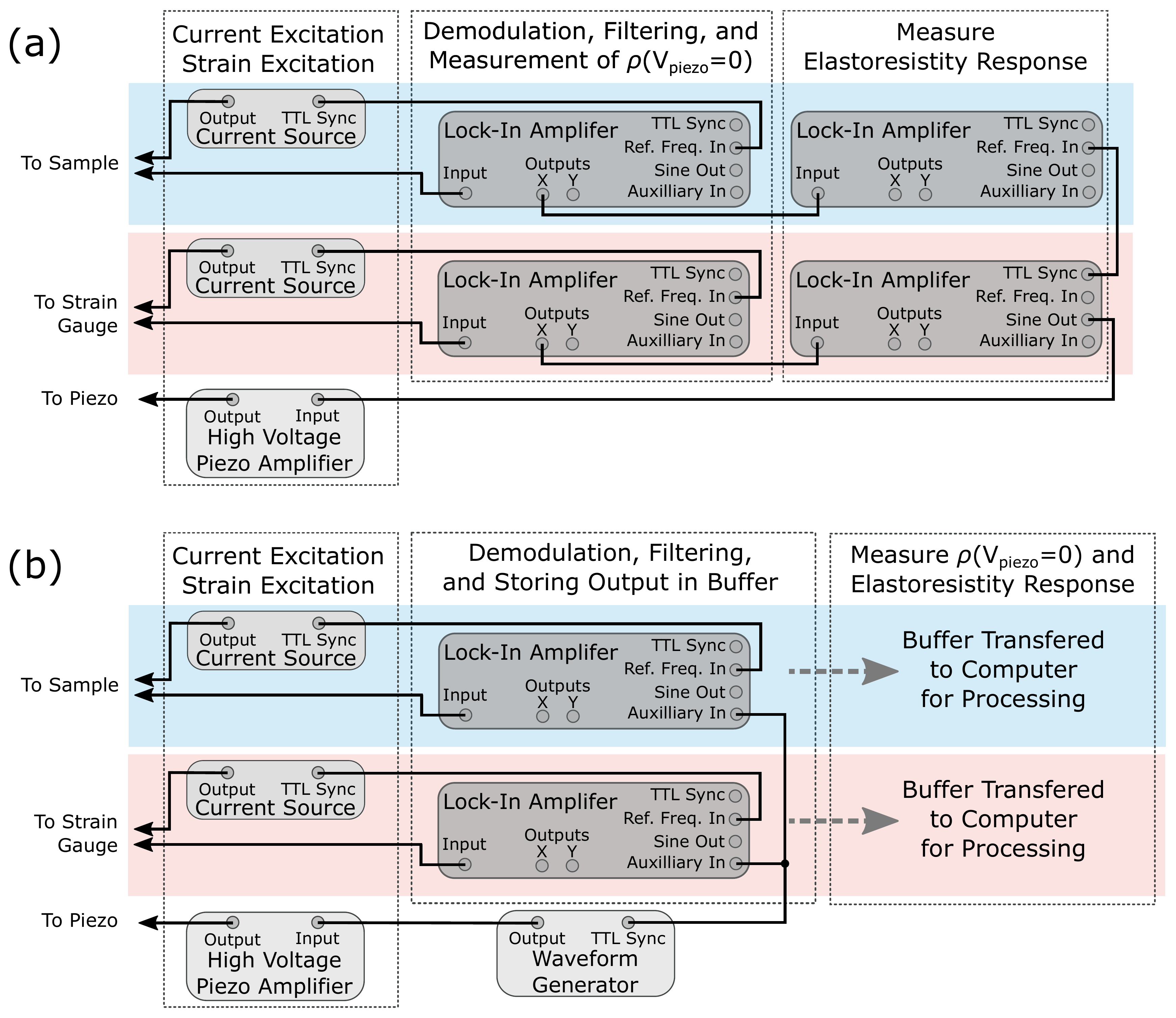}
\caption{Wiring diagram for two AC elastoresistivity detection methods. 
All lock-in amplifiers are connected to a computer for data acquisition and synchronization to temperature control (not pictured).
Their ``X'' and ``Y" outputs provide the in-phase and in-quadrature components of the demodulated signal, respectively. 
(a) The ``dual lock-in method" described in \cref{sec:dll} uses two lock-in amplifiers connected in series. 
The first of these measures the unstrained resistance, and the second detects the amplitude modulation of the resistance.
The second lock-in for each channel is phase locked to the voltage input to the piezo amplifier, which enables detection of relative phase between the elastoresistivity response and the strain.  
(b) The ``buffered acquisition method'' described in \cref{sec:bam} uses a single lock-in amplifier per channel. 
The sample resistivity is measured over time by recording the lock-in output, up to the maximum buffer acqusition frequency of the instrument. 
Post processing extracts the component of resistivity varying at the same frequency as the strain $\omega_s$.
The relative phase of strain and elastoresistivity is determined by comparing each to a measured reference signal from the function generator driving the piezo amplifier; the reference can also be recorded to the buffer through auxiliary inputs.
}
\label{fig:dual}
\end{center}
\end{figure*}

This section presents methods for detecting the elastoresistivity signal described in \cref{theory}.
Rather than directly measuring the voltage at one of the sideband frequencies, we implement a demodulation technique to transform the elastoresistivity signal at the sideband frequencies $\omega_c\pm\omega_s$ into a signal at the strain frequency $\omega_s$, which is then measured.
Signal acquisition for this task can be considered in three separate stages: demodulation, filtering, and detection.
The first two of these three steps are implemented inside a digital lock-in amplifier (Stanford Research Systems 830), which fundamentally comprises a mixer and low-pass filter. 
It should be emphasized, however, that the method described in\cref{theory} and shown in \cref{fig:schematic} is independent of this particular choice of hardware.

\subsubsection{Use of Lock-In Amplifier as Demodulator and Filter}

The mixing components of a lock-in amplifier multiply the voltage signal from a sample (strain gauge) against a reference oscillation, which is at the same frequency as the current excitation on the sample (strain gauge).
The resulting product signal has components at the sum and difference frequencies of the input and reference signals. 
The components at the difference frequencies comprise the elastoresistivity signal at $\omega_s$ and the unmodulated resistivity signal as a DC voltage, while the sum, or $2f$ component, occurs at approximately twice the carrier frequency.
A lock-in amplifier removes the high frequency component with a built-in low-pass filter; if a lock-in is used to perform the demodulation, the filter may also attenuate the elastoresistive signal at $\omega_s$ if care is not taken to select appropriate frequencies and filter parameters. 


The effect of a low pass filter in a lock-in amplifier can be characterized by a transfer function $T(\omega)$.
In the elastoresistivity experiment performed here, there are two sidebands at frequency $\omega_c\pm\omega_s$.
Treating the amplitudes of each of these as complex values, the output from the lock-in is given by
\begin{equation}
\tilde{V}_{out}(\omega_s) = T(\omega_s)(\tilde{V}_{in}(\omega_c+\omega_s)+\tilde{V}_{in}(\omega_c-\omega_s)),
\end{equation}
so the sideband elastoresistivity signals are obtained from the output of the lock-in according to
\begin{equation}
\tilde{V}_\pm\equiv \tilde{V}(\omega_c+\omega_s)+\tilde{V}(\omega_c-\omega_s) =\frac{\tilde{V}_{out}(\omega_s)}{T(\omega_s)},
\label{eq:Demodulated}
\end{equation}
where $\tilde{V}(\omega)$ is the Fourier component of the voltage across the sample at frequency $\omega$, which is related to the resistivity and elastoresistivity of the sample according to \cref{eq:SampleSidebands}.

\subsubsection{Dual Lock-in Method}
\label{sec:dll}
The demodulated elastoresistance signal in \cref{eq:Demodulated} can be detected using a second lock-in amplifier, referenced to the signal generator providing the voltage to the piezo-amplifier at strain frequency $\omega_s$.
The electronics for measuring a single resistance from a sample using this implementation is shown in \cref{fig:dual} (a).
In this configuration, the electrical outputs of the first lock-in amplifier also provide a gain to the input voltage $G_\mu$ where $\mu$ is the full-scale sensitivity setting of the instrument (measured in volts).
For the model of amplifier used here, $G_\mu =$ ($10$~\volt)/$\mu$.
The value recorded by the second lock-in amplifier must therefore be divided by $G_\mu T(\omega_s)$ to obtain the actual voltage on the sample.
This setup can be repeated on a resistive strain gauge to measure the strain experienced by the sample, which is necessary to extract the elastoresistivity response.

\subsubsection{Buffered Acquisition Method}
\label{sec:bam}
The major drawback of the method described above is the requirement of two lock-in amplifiers per measurement channel. 
Full in-plane symmetry decomposition of the elastoresistivity response requires at least two resistivity measurements and at least one strain measurement;\cite{Shapiro_Measurement_2016} the technique above quickly becomes impractical if multiple strain gauges or samples are measured.
A simple modification to this method reduces the number of instruments needed: rather than sending the output of the lock-in amplifier to a second instrument, the output can be sampled and stored in a buffer internal to the instrument and then transferred to a computer for post-processing to extract the $\omega_s$ component.
To extract the relative phase of the strain gauge and sample resistivities, the lock-in amplifiers can be synchronized to a reference TTL signal which is recorded through the auxiliary inputs, as shown in \cref{fig:dual} (b).
The maximum strain frequency which can be used in this technique is determined by the sampling rate of the lock-in amplifier rather than the maximum internal reference frequency of the lock-in amplifier.
Thus, this technique trades instrumental complexity for maximum strain frequency. 
Even in the quasi-DC limit of a few hertz, however, this method still provides a significant speedup compared to the traditional method of stepping the piezoelectric, and no longer carries the stringent temperature stability requirements.
Furthermore, more recent models of lock-in amplifier have significantly higher maximum frequency in this mode.

\section{Measurement Errors} \label{error}

\subsection{Factors common to DC and AC elastoresistivity}

Because the DC and AC methods for measuring elastoresistivity described here can be used in similar geometries and with identical techniques for mounting crystals to piezoelectric devices, there are many sources of error that are common to both. 
These include factors affecting the mixing of elastoresistivity coefficients in a measurement: the sample contacts may be misaligned relative to each other or relative to the sample axes, and the crystal may be mis-oriented on the the piezoelectric stack.
Furthermore, strain to the sample may be inhomogeneous and a large offset strain might be induced by differential thermal contraction of the piezoelectric and the sample when the sample is directly mounted to the piezoelectric stack.
In these respects, the errors introduced to the measurement are identical, and so we refer readers to previous analyses of the DC elastoresistivity technique.\cite{Shapiro_Measurement_2016}

\subsection{The Effects of Temperature Instability on AC and DC Elastoresistivity Measurements}
Temperature drifts in a cryostat can introduce significant errors to the measurement of the elastoresistivity.
If the resistivity or elastoresistivity of the sample is temperature dependent, then the measured resistivity change in a strained sample is approximated by
\begin{eqnarray}
\Delta\rho_{ij}(T,\varepsilon)& \approx & \rnorm_{ij}(T) \sum_{kl} \left(m_{ijkl}(T_0)\varepsilon_{kl} \right) \nonumber
\\
& +& \left(\frac{\partial \rho_{ij}}{\partial T}+\rnorm_{ij}(T) \sum_{kl} \frac{\partial m_{ijkl}}{\partial T}\varepsilon_{kl} \right)\delta T,\label{eq:TempInstability}\nonumber
\\
\end{eqnarray}
which shows that temperature fluctuations introduce additional time-dependence in the resistivity that can obscure the elastoresistivity signal.
For this reason, we have found that a DC elastoresistivity measurement must be performed when the cryostat has completely thermally equilibrated; even the decaying temperature oscillations from a PID temperature controller can obscure the elastoresistivity of a sample.
This precaution considerably adds to the time required to perform a DC elastoresistivity measurement.

In contrast, the AC elastoresistivity technique performs a demodulation measurement to obtain only the $\omega_s$ component of \cref{eq:TempInstability} and is therefore only sensitive to the $\omega_s$ component of $\delta T$ in the regime of strains used, ranging here from 10~Hz to 3kHz.
Temperature variation can occur at frequency $\omega_s$ as either (1) a component of overall temperature fluctuations of the cryostat, or (2) a result of endogenous heating from the elastoresistivity experiment.

For the first of these two effects, cryostats can be designed to attenuate the $\omega_s$ component of temperature fluctuations: the specific heat of the sample stage and the thermal conductivity between sample stage and heat exchanger can be adjusted to form a thermal low-pass filter. 
In fact, we have been able to use the AC elastoresistivity technique even for strain frequencies as low as 3~Hz as the cryostat temperature is swept continuously at 1~K/min.

For heating originating at the experiment, we are concerned primarily with elastocaloric heating of the piezoelectric at the same frequency as the strain.
While other heating effects are known to us (see \cref{app:Rxx}), these occur predominantly at either DC or second harmonics of the strain frequency and are therefore filtered out by the measurement electronics.
In the DC elastoresistivity experiment, elastocaloric effects can be safely neglected because the sample resistivity is measured only after the sample has had sufficient time to thermalize through its leads following a change in strain.
For strain frequencies greater than 1~kHz, elastocaloric heating of the sample from the piezoelectric must also vanish; the thermal penetration depth for a polymer such as the Devcon 5-Minute Epoxy used in \cref{demo} becomes less than 100~nm for strain frequencies greater than 1kHz, which is significantly less than the thickness of the glue estimated in Ref. \onlinecite{Hicks_Piezoelectric_2014}.
In the intermediate regime, there is no observable difference of the measured elastoresistivity from either the DC or high frequency regime, suggesting that elastocaloric effects can be safely neglected.
Ultimately, elastocaloric effects can also be attenuated by mounting the sample away from the surface of the piezoelectric stack, as has previously been done for DC strain.\cite{Hicks_Piezoelectric_2014}

\section{Measurement of Elastoresistivity of a Fe-based Superconductor} \label{demo}

As a demonstration of this technique, we present a measurement of a prototypical underdoped iron pnictide, BaFe$_{1.975}$Co$_{0.025}$As$_2$. 
This material has an electronically driven tetragonal-to-orthorhombic structural transition at $T_s=$98~K.\cite{Chu_Determination_2009} 
At temperatures above the structural transition, $2m_{xyxy}$ relates to the thermodynamic susceptibility of the order parameter.\cite{Shapiro_Symmetry_2015} 
Consequently, $m_{xyxy}(\omega)$ follows a Curie-Weiss law in the $\omega\to0$ limit in the absence of significant disorder or dissipative effects.\cite{Kuo_Ubiquitous_2016}
The thermodynamic significance of $m_{xyxy}$ motivated measurements which isolate this individual component of the elastoresistivity tensor. 
Here we show that the present AC demodulation technique detects this diverging electronic response to strain, reproduces prior DC measurement techniques in the quasi-static strain limit, and performs better in the quasistatic strain limit than previous DC elastoresistivity measurements.

\begin{figure} 
\begin{center}  
\includegraphics[width=\columnwidth]{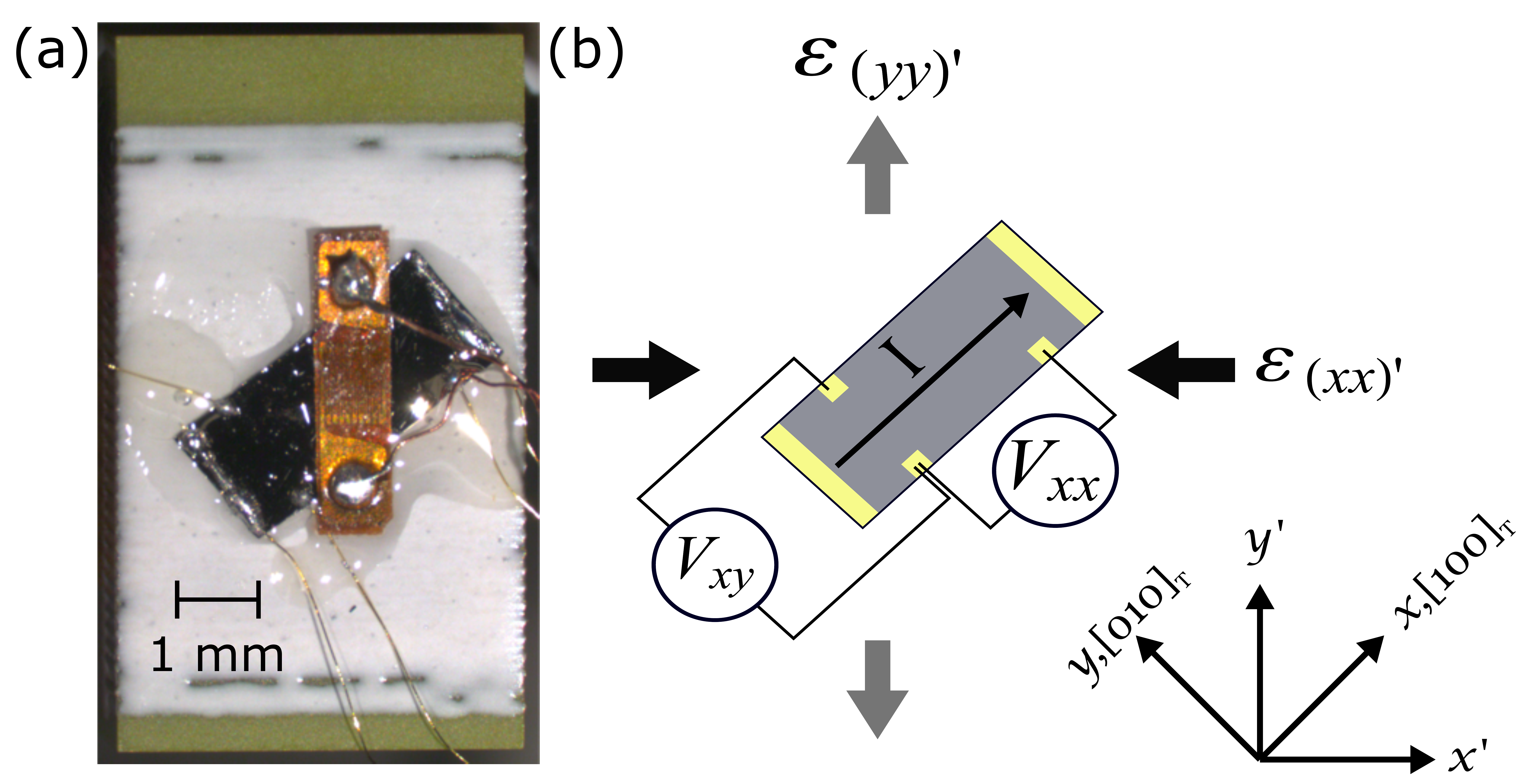}
\caption{\small \sl (a) Image of the crystal glued to a piezoelectric stack in a ``transverse" configuration, with current along the axes of the tetragonal primitive unit cell. The piezoelectric stack shown is 5mm wide. A strain gauge was affixed to the surface in order to test strain transmission as detailed in \cref{app:strain}.  (b) Electrical schematic showing the distance between contacts and the relative alignments of contacts, the crystal , and the axes of the piezoelectric axes.}
\label{fig:demo_config} 
\end{center}  
\end{figure}

Isolating individual elastoresistivity components depends on the crystal structure.
For a tetragonal material like \sampleNS,
\begin{equation}
\left(\frac{\Delta\rho}{\rho_0}\right)_{xy} =2m_{xyxy}\varepsilon_{xy},
\end{equation}
and symmetry guarantees $(\Delta\rho/\rho_0)_{xy}$ is independent of all other strain at linear order.
Furthermore, for this crystal structure,\cite{Shapiro_Symmetry_2015}
\begin{equation}
\left(\frac{\Delta\rho}{\rho_0}\right)_{xy}=\frac{\Delta\rho_{xy}(\varepsilon)}{\rho_{xx}(\varepsilon=0)},
\end{equation}
such that, 
\begin{align}
2m_{xyxy} &=
\frac{1}{\varepsilon_{xy}}
\frac{{\Delta\rho_{xy}}(\varepsilon_{xy})}{\rho_{xx}(\varepsilon=0)}.
\label{eq:2mxymyratio}
\end{align}
In both the DC and AC elastoresistivity measurements, we use the resistivity measured with the piezo stack terminals shorted together as a proxy for $\rho_{xx}(\varepsilon=0)$. 
In practice, this assumption can be avoided: the sample can either be mounted on an apparatus that cancels the thermal expansion of the active piezoelectric material,\cite{Hicks_Piezoelectric_2014} or a DC bias voltage can be applied to the piezoelectric stack to reduce the effects of differential thermal contraction.

The sample was grown by a self-flux method as described elsewhere, \cite{Yamamoto_Small_2009,Chu_Determination_2009} and then cut into a rectangular bar with edges along the [100] and [010] tetragonal crystallographic axes ( 3701~\micro\meter ~long, 1728~\micro\meter ~wide and 20~\micro\meter ~thick).
Electrical contact to the sample was made with gold wires through Chipquik SMD291AX10T5 solder.
The contacts were arranged in a transverse configuration, as is detailed in Ref.\onlinecite{Shapiro_Measurement_2016}.
The longitudinal contacts were estimated to be separated by approximately 2400~\micro\meter.
Current was sourced along the long axis of the rectangular sample, and voltage contacts were positioned to directly measure $\rho_{xy}$ and $\rho_{xx}$
\footnote{For this geometry and material, the elastoresistivity response along the current direction is approximately 10 times smaller than the response perpendicular to the current. 
Furthermore, as the contact contamination for the resistivity was found to be less than a few percent for this sample, there was no need to subtract the contamination of the longitudinal elastoresistivity, though a thorough procedure is detailed in Ref. \onlinecite{Shapiro_Measurement_2016}.}.
The crystal was affixed to the Piezomechanik PSt150/5x5/7 cryo 1 piezoelectric stack with Devcon 5-Minute Epoxy. 
In so mounting the crystal, the crystallographic axes were rotated by $45^\circ$ with respect to the axes of the piezoelectric, as seen in \cref{fig:demo_config}. 
The experiment was carried out in an Oxford Instruments OptistatCF cryostat with the sample in static exchange gas. 

Strain along the axes of the piezoelectric stack was measured using a bidirectional linear resistive strain gauge (Part No. WK-05-062TT-350-L from Vishay Precision Group). 
Strain transmission through the sample was measured by a unidirectional strain gauge (Part No. WK-05-031DE-350) glued onto the surface of the sample and oriented along the $y'$ axes of the piezoelectric. 
Because the sample was rotated relative to the axes of the piezoelectric stack, the shear strain experienced by the sample is related to the anisotropic strain of the piezoelectric stack $\varepsilon_{xy}=-(\varepsilon_{x'x'}-\varepsilon_{y'y'})/2$. 
This expression can be further simplified to extract the shear strain experienced by the sample from a single strain gauge, as the piezoelectric Poisson ratio, $\nu_P = -\varepsilon_{y'y'} /\varepsilon_{x'x'}$, was previously characterized.

The sample current was sourced from a Keithley 6221 AC current source at $\omega_c\approx$14~kHz  with an amplitude of 5~mA rms.
Each strain gauge was wired into a separate Wheatstone bridge circuit, which was driven with an AC current of 1~mA rms and a distinct frequency of order 10~kHz.
The differential voltage output of the sample and strain gauge circuits were measured by a Stanford Research 830 lock-in amplifier.

AC strain was induced in the sample by driving the piezoelectric stack with a 25~V peak-to-peak amplitude sine wave with frequency $\omega_s$ ranging from 10~Hz to 3~kHz. 
This voltage was produced by amplifying the sine-out signal from a Stanford Research 830 lock-in amplifier with a Tegam 2350 high-voltage amplifier. 
The output of the voltage amplifier was monitored by the lock-in amplifier through the Tegam 2350's included 1:100 buffered voltage divider.

The elastoresistivity as a function of strain frequency and temperature was obtained by stabilizing at a set of fixed temperatures between 100~K and 240~K.
At each temperature, the frequency of the voltage excitation to the piezoelectric stack was incremented on a logarithmic scale between 10~Hz and 3~kHz.

\begin{figure}[!ht]  
\begin{center}  
\includegraphics[width=\columnwidth]{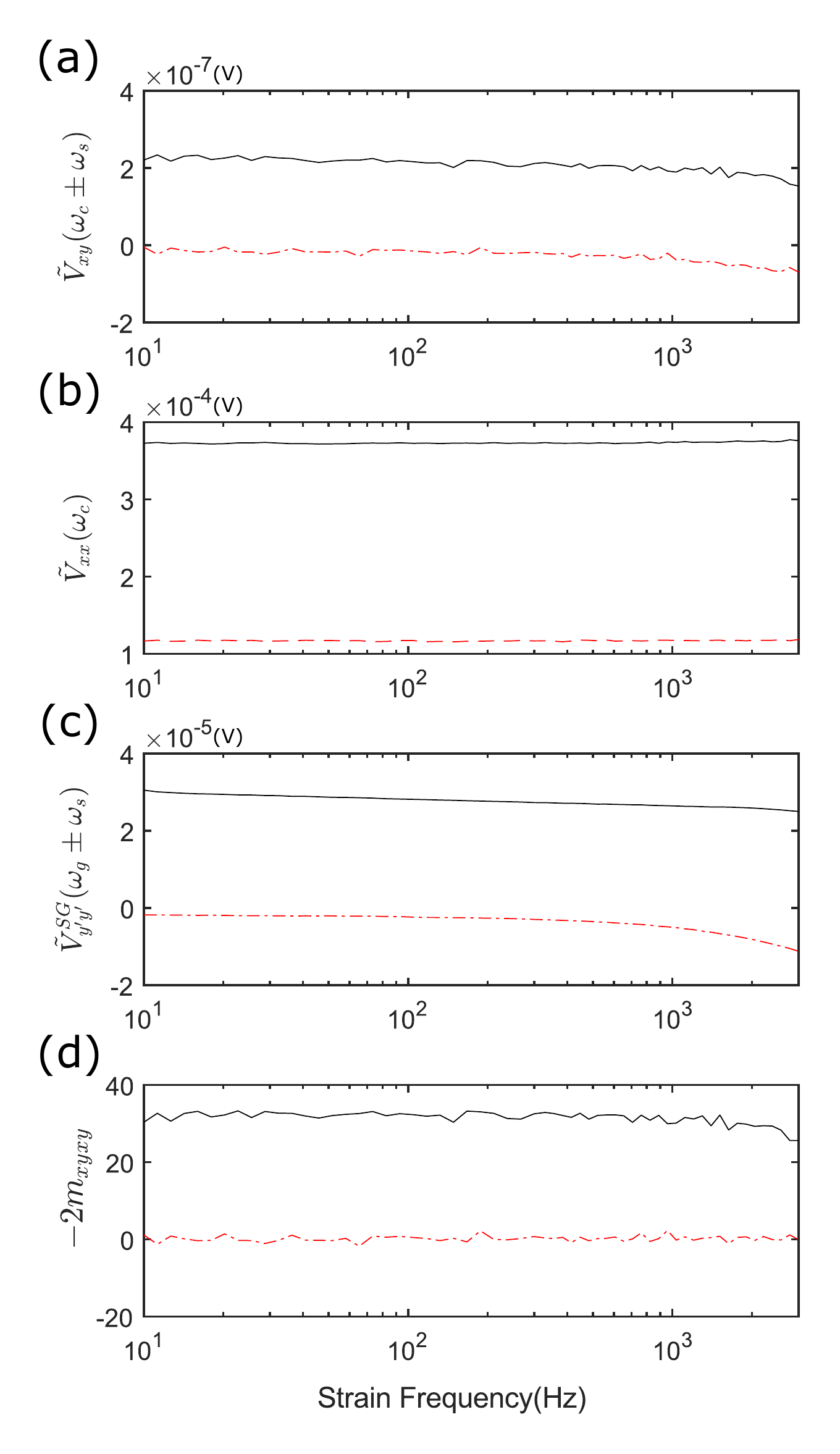}
\caption{\small \sl Representative data for an elastoresistivity measurement of \sample taken at 133K showing the real(black) and imaginary(red) voltages obtained for various strain frequencies, $f_s=\omega_s/2\pi$, used in an elastoresistivity measurement. Values at sideband voltages are obtained by dividing out the gain and filtering functions of the lock-in amplifier, as described in \cref{app:TransferFunction}. (a)  Voltage component across the transverse contacts of the samples, measured from the sum of amplitudes of the sideband frequencies $\omega_c\pm\omega_s$.   (b) Voltage across the longitudinal contacts of the sample at the carrier frequency $\omega_c$ is proportional to the average resistance of the sample, and is used to obtain the normalization value $\rnorm_{xy}$. 
(c) Sum of voltage amplitudes at the two sideband frequencies  $\omega_g\pm\omega_s$ is proportional to the strain on the sample. 
(d) The calculated value of $m_{xyxy}$ obtained from \cref{eq:m66calc}. 
The downturn above 1~kHz results from heating of the piezoelectric stack, as discussed in \cref{app:Rxx}.}
\label{fig:trace}  
\end{center}  
\end{figure}

In this experiment $m_{xyxy}$ was computed by isolating each of the three quantities on the right side of \cref{eq:2mxymyratio}. 
Including all of the various geometric factors, these quantities of interest were obtained from their relation to the following measured voltages:
\begin{align}
\tilde{V}_{xx}( \omega_c) &=  I_0\frac{L}{WH}\rho_{xx}(\varepsilon=0) \\
\tilde{V}_{xy} (\omega_{\pm})&= \frac{1}{2}\frac{I_0}{H}\rho_{xx}(\varepsilon=0) 2m_{xyxy}\varepsilon_{xy,0} \\
\tilde{V}^{SG}_{y'y'}(\omega_{g\pm}) &= \frac{1}{2}I^{\text{SG}}_0\frac{dR^{\text{SG}}}{d\varepsilon_{y'y'}}\varepsilon_{y'y',0} \nonumber \\
&= \frac{1}{2}I^{\text{SG}}_0\frac{dR^{\text{SG}}}{d\varepsilon_{y'y'}}\left(\frac{2\nu_P}{\nu_P+1}\right)\varepsilon_{xy,0},
\end{align}
where $\varepsilon_{xy,0}$ is the amplitude of the induced strain oscillation at angular frequency $\omega_s$, $I_0$ is the current in the sample, $I^{\text{SG}}_0$ is the current to the strain gauge, $L$ is the length between longitudinal contacts,  $W$ is the sample width and $H$ is the sample thickness.
The derivation is detailed in \cref{app:Poisson}.
From this we can calculate the $2m_{xyxy}$ elastoresistivity response using this AC method,
\begin{align} \label{eq:m66calc}
2m_{xyxy}=I^{SG}_{0}\frac{L}{W}\frac{dR^{SG}}{d\varepsilon_{y'y'}} \left(\frac{2\nu_P}{\nu_P+1}\right)\frac{V_{xy,\omega_\pm}}{V_{xx, \omega_c}V^{SG}_{y'y', \omega_\pm}}.
\end{align}

A representative data trace taken at 133~K using the dual lock-in technique described in \cref{sec:dll} is shown in \cref{fig:trace}.
For these calculations we used the measured DC Poisson ratio of the piezoelectric stack to calculate $\varepsilon_{xy}$ from a single strain gauge measurement (See \cref{app:Poisson}). 
We also have ensured that approximately 100\% of the strain on the piezoelectric stack is transmitted through the sample, as shown in \cref{app:strain}.

For comparison, measurements were also made using the DC elastoresistivity technique in the transverse configuration.\cite{Shapiro_Measurement_2016}
The sample was stabilized at a range of similar temperatures, and strain was swept by varying the voltage on the piezoelectric in 5~V increments between -50~V and 50~V.
The resistivity was recorded at each voltage, and a linear fit at each temperature was used to obtain the DC value of $m_{xyxy}$.

\begin{figure}[t]  
\begin{center}  
\includegraphics[width=\columnwidth]{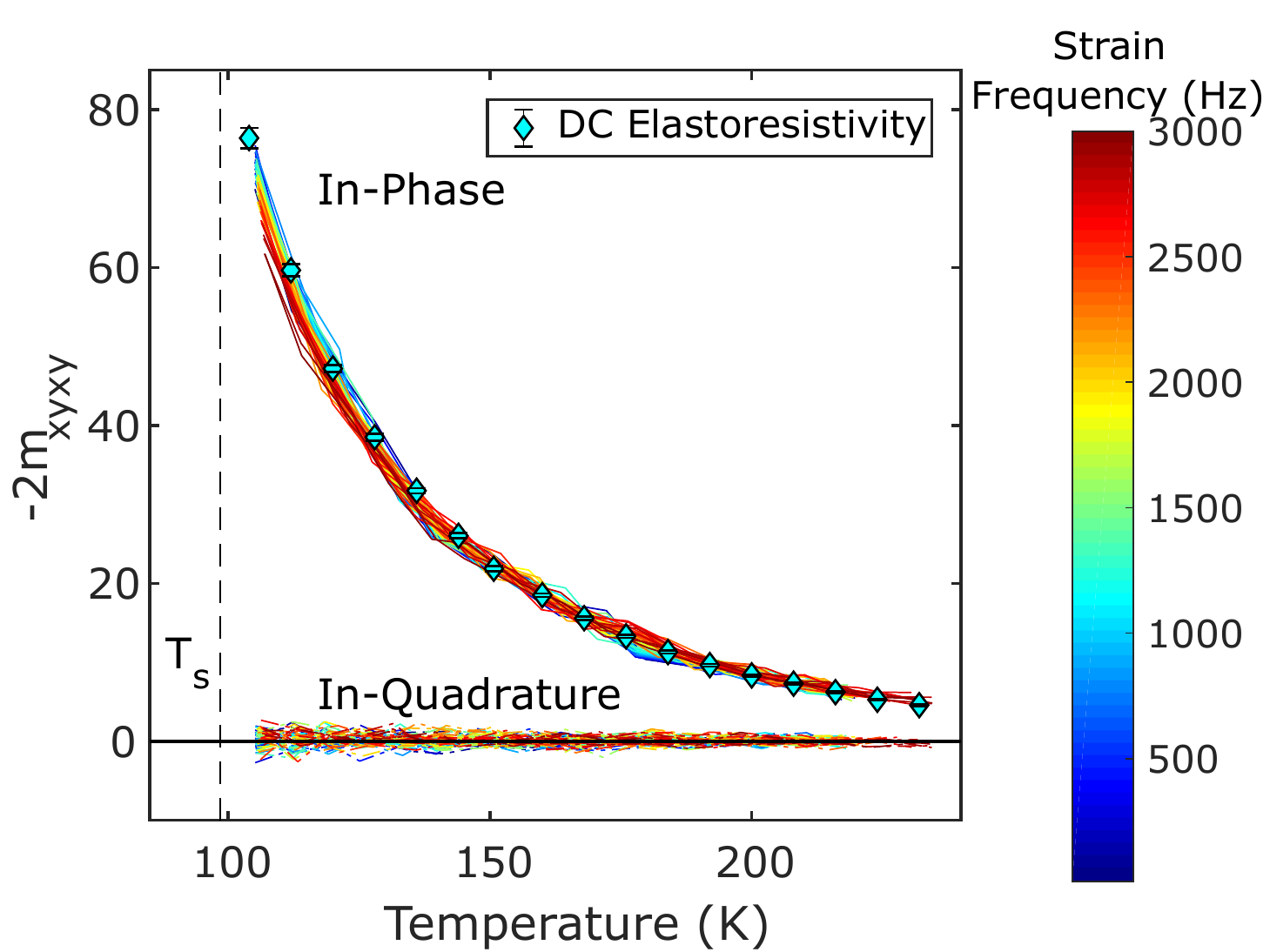}
\caption{\small \sl The elastoresistivity coefficient $m_{xyxy}$ of \sample as function of frequency and temperature. 
Overlaid is a DC trace on the same sample (blue diamonds).
}
\label{fig:m66}  
\end{center}  
\end{figure}

The nematic susceptibility at finite frequency is necessarily a complex quantity.
The elastoresistivity must therefore be described in terms of an in-phase and in-quadrature response, which can be obtained naturally from \cref{eq:m66calc} by treating the lock-in voltages as complex quantities.
The results of this calculation over a full range of temperatures (105~K to 225~K) and frequencies (3~Hz to 3~kHz) are shown in \cref{fig:m66}.
When the effects of piezoelectric heating are taken into account (by measurement in \cref{app:Rxx}) the extracted value of $2m_{xyxy}$ collapses onto the value measured by the DC method at low frequencies.
The quantitative agreement between these two methods demonstrate the robustness of this technique. 
Furthermore, these measurements also show that, over the frequencies and temperatures measured here, there is no significant frequency dependence in $m_{xyxy}$.

\begin{figure}[t]  
\begin{center}  
\includegraphics[width=\columnwidth]{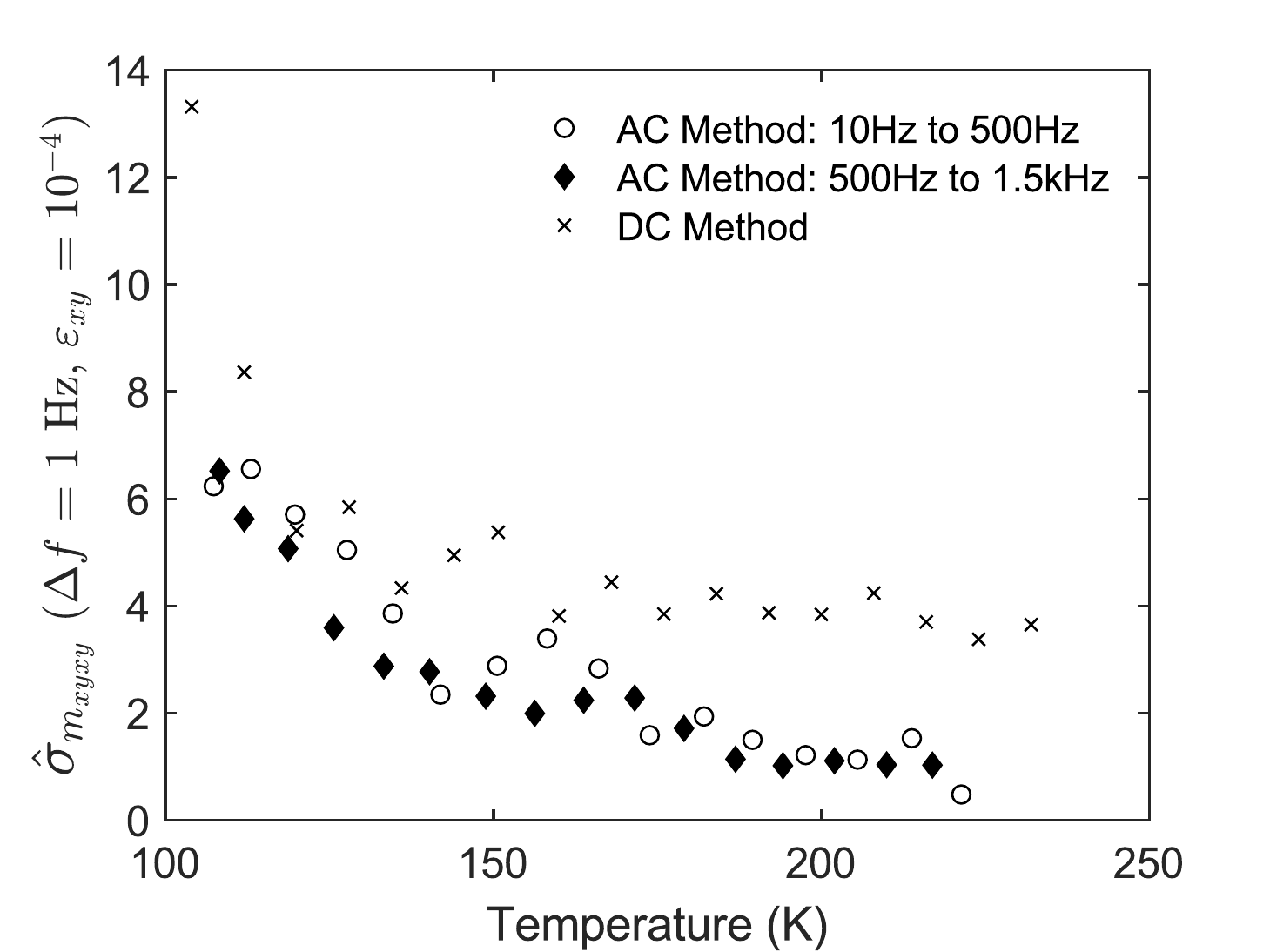}
\caption{\small \sl Noise of the elastoresistivity measurement performed by the DC and AC techniques, standardized for a measurement with 100ppm strain and a 1~Hz effective noise bandwidth. 
The AC technique performs similarly in both the 10-500~Hz and 500-1500~Hz regimes, and appears to perform better than the DC technique.}
\label{fig:SNR}  
\end{center}  
\end{figure}

We find it instructive to compare not only the measured values of the elastoresistivity, but also the noise in these measurements. 
Though \cref{fig:m66} shows greater variability in the individual measurements of $m_{xyxy}$ acquired with an AC technique compared to those acquired with a DC technique, each AC elastoresistivity measurement is acquired with four times less voltage on the piezoelectric stack and in approximately one tenth the time.
To accurately capture this discrepancy in measurement conditions, the standard error of the measurement, $\hat{\sigma}_{m_{xyxy}}$, must be scaled appropriately.
For the DC elastoresistivity, the regression which was used to obtain $m_{xyxy}$ also provides a lower bound
\footnote{Simple linear regression assumes independent errors, but since temperature fluctuations are introduce correlations, the standard error obtained from regression is a lower bound.} 
of $\hat{\sigma}_{m_{xyxy}}$.
For an estimate of the noise in the AC elastoresistivity measurement, an upper bound
\footnote{
We assume no frequency dependence of the elastoresistivity in order to compute the variance of the measured $m_{xyxy}$ values. Any present frequency dependence will increase the variance and provide an overestimate.
}
of $\hat{\sigma}_{m_{xyxy}}$ is obtained from the variance of measured elastoresistivities at a given temperature in the frequency regime between 10~Hz and 1.5~kHz.
These estimates of $\hat{\sigma}_{m_{xyxy}}$ must then be standardized for equal bandwidth and equal strain.
Since no closed form is known to us for the effective bandwidth of a linear regression, we used the full time between independent measurements as a proxy for bandwidth for the DC technique. 
The strain was standardized to 100 ppm, which is approximately what is obtained for a 25\volt peak-to-peak excitation on the piezoelectric stack.
The measurement noise for a $\varepsilon_{xy}=10^{-4}$ strain with an equivalent noise bandwith of 1~Hz is shown in \cref{fig:SNR}.
The effective noise in the measurement of $m_{xyxy}$ was less for the AC method by a factor of approximately 3 over much of the temperature regime, which means that the AC elastoresistivity technique can be performed faster or with less induced strain on the sample.
Close to the transition, the noise is found to increase, possibly due to the increase in $dm_{ijkl}/dT$, but this increase is found to be slower than the corresponding increase in $m_{xyxy}$, so the signal to noise ratio increases as $m_{xyxy}$ increases.

\section{Conclusion}

The technique described in this paper enables measurement of the dynamical elastoresistivity $m_{ijkl}(\omega)$, a physical quantity that has previously not been considered beyond the zero frequency limit.
These measurements provide access to a regime of slow dynamics that is otherwise inaccessible by other standard methods. 
In particular, since certain terms in the elastoresistivity tensor are proportional to the nematic susceptibility of a material,\cite{Shapiro_Symmetry_2015} measurements of the dynamic elastoresistivity reveal the frequency dependence of the nematic susceptibility in a very different frequency regime to those probed by other standard techniques.
For example, nematic susceptibility measurements have also been performed with Raman scattering\cite{Gallais_Observation_2013, Massat_Charge_2016, Thorm_Critical_2016} and resonant ultrasound spectroscopy, \cite{Migliori_Implementation_2005, Fernandes_Effects_2010} which typically measure in regimes above 1~GHz and 300~kHz, respectively. 
Furthermore, since this technique is based on electronic transport, elastoresistivity is especially sensitive to dynamics affecting quasiparticles at the Fermi energy, and can be measured with very high precision.

Classes of materials that can be experimentally investigated with this technique might include those exhibiting either nematic glass behavior or those realizing a random-field Ising nematic system due to the presence of disorder.
Indeed, the operating frequency range of this AC elastoresistivity technique spans much of the frequency regime used in AC magnetic susceptibility measurements to determine the existence of activated behavior in magnetic random-field Ising systems.\cite{Nash_Experimental_1991}

This technique can also be used in the low frequency limit to extract the same physical quantity as the DC elastoresistivity, but with several significant advantages.
By using dynamic strain and lock-in techniques, this technique can operate with lower elastoresistivity signals and lower voltages for driving piezoelectric devices, reducing mechanical wear on the sample and electrical contacts.
Additionally, the smaller timescales required by this technique accelerates data acquisition and opens up new measurement regimes; one such promising regime includes high magnetic fields accessible only for short times in pulsed magnets.

\begin{acknowledgments}
We acknowledge helpful conversations with M. Ikeda and P. Walmsley.
A.T.H., J.C.P., and T.A.M. are supported by a NSF Graduate Research Fellowship under grant DGE-114747.
J.C.P. is also supported by a Gabilan Stanford Graduate Fellowship.
J.S. acknowledges support as an ABB Stanford Graduate Fellow.
This work was supported by the Department of Energy, Office of Basic Energy Sciences, under contract no. DE-AC02-76SF00515.
\end{acknowledgments}


\begin{appendices}
\crefalias{section}{appsec}
\crefalias{subsection}{appsec}


\section{Strain-per-Volt Characteristics of Piezoelectric Stacks}
\label{app:strainpervolt}
\begin{figure}[!ht]  
\begin{center}  
\includegraphics[width=\columnwidth]{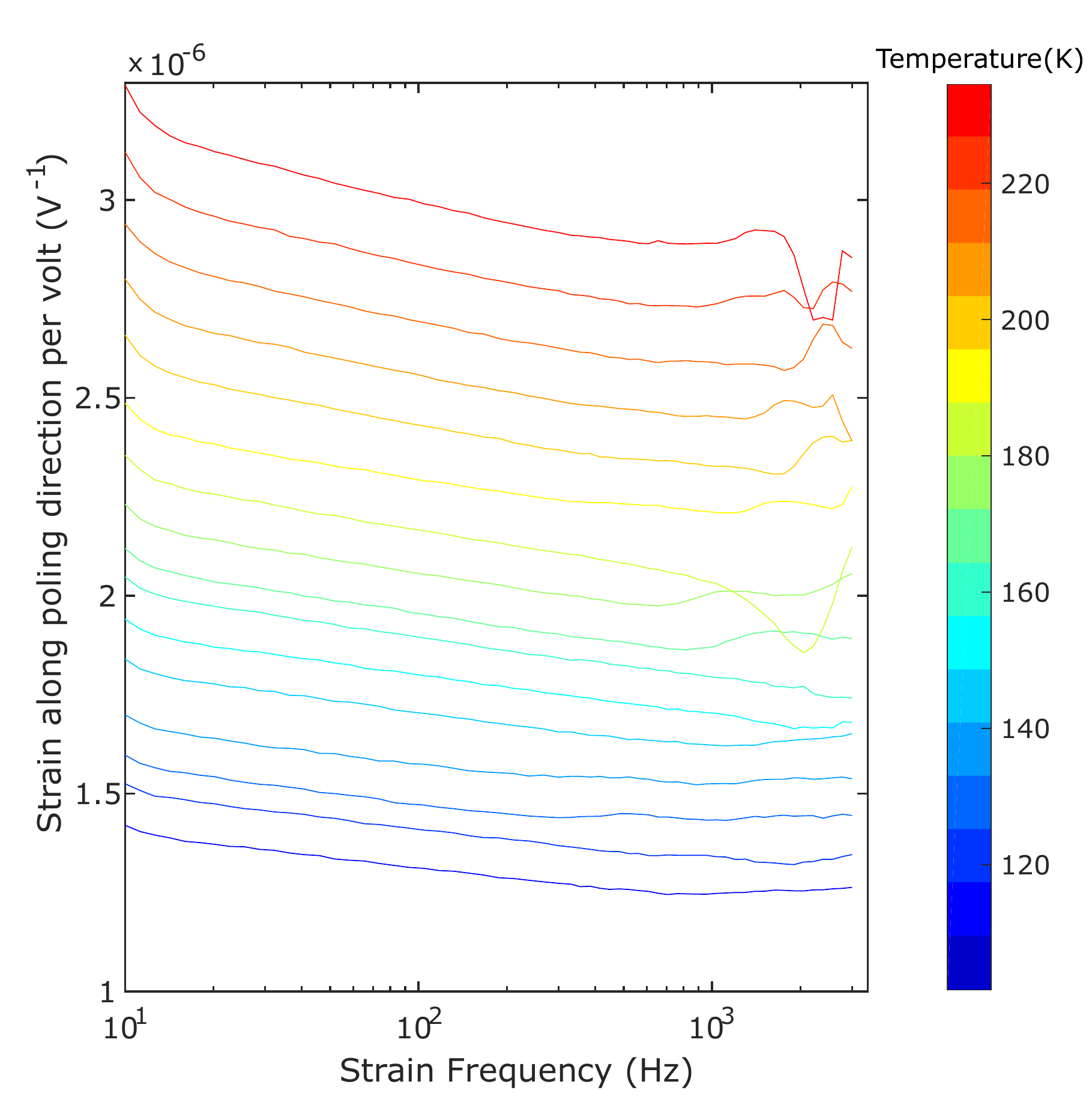}
\caption{\small \sl Strain along poling direction of piezoelectric stack per unit voltage on the piezoelectric stack used in this study, as measured by a resististive strain gauge adhered to the surface of the stack with Devcon 5-Minute Epoxy.
Excepting the data at high temperature and high frequency, which were acquired last and which may be affected by aging of the glue, the data reveal a monotonic dependence on frequency and temperature.
\label{fig:PZT_StrainPerVolt}}
\end{center}  
\end{figure}

The electromechanical response of the Piezomechanik ``PSt150/5x5/7 cryo 1'' was measured by mounting a resistive strain gauge with Devcon 5-Minute Epoxy along the long axis of the piezo and is shown in \cref{fig:PZT_StrainPerVolt}. 
The techniques outlined in \cref{theory} were used to measure the strain, while the voltage on the piezo stack was measured through a high impedance 1:100 voltage divider.
Both strain and voltage must be measured with lock-in methods to isolate individual frequencies in the case of harmonic distortion from the amplifier.
The strain induced in the strain gauge decreases significantly with decreasing temperature, and  decreases with frequency as well. 
Despite the diminution of induced strain at low temperatures and high frequencies, the strain induced is still sufficient to accurately measure an elastoresistive response.

\section{Temperature Dependent Capacitance of a Piezoelectric Stack}
\label{app:Capacitance}
\begin{figure}[!ht]  
\begin{center}  
\includegraphics[width=\columnwidth]{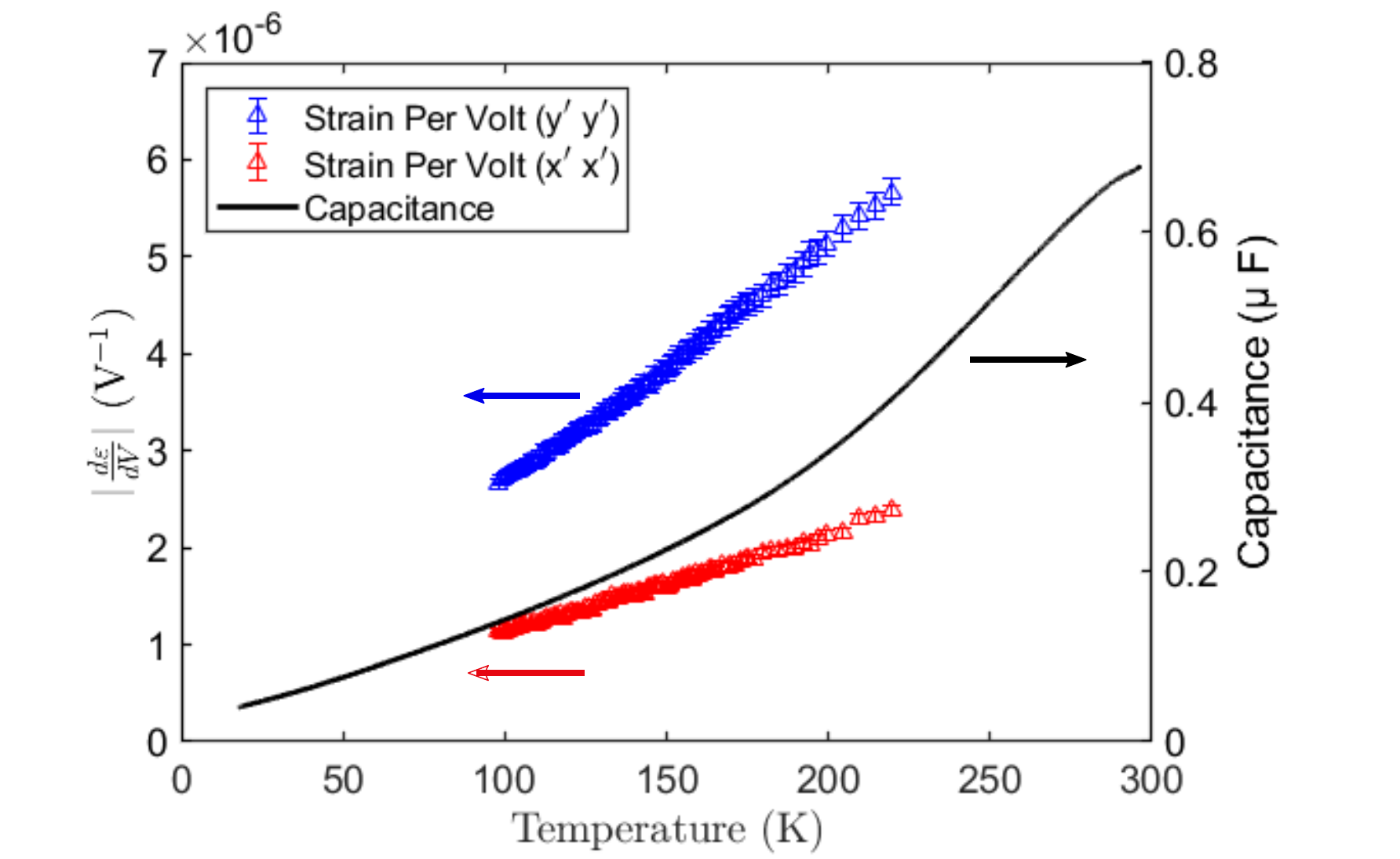}
\caption{\small \sl Temperature dependence of the capacitance of a Piezomechanik ``PSt150/5x5/7 cryo 1'' stack (right axis) and the strain along the poling and perpendicular direction (left axis) over the regime of temperatures measured in this work. Details of the specifice piezoelectric stack used in this study are given in the main text.
This characterization can be different for different models of piezoelectric stack.
\label{fig:PZT_Capacitance}  }
\end{center}  
\end{figure}

The significant capacitance of a piezoelectric stack determines the electrical current necessary  to tune strain at higher frequencies and amplitudes.
The capacitance of the Piezomechanik ``PSt150/5x5/7 cryo 1'' piezoelectric stack was measured using an Andeen-Hagerling AH2550 1~kHz capacitance bridge, and is shown in \cref{fig:PZT_Capacitance}.
The capacitance decreases by roughly an order of magnitude between room temperature and 20 K, and decreases upon cooling faster than the strain per volt of the piezoelectric stack, making the stack more suitable for high frequency applications at cryogenic temperatures.

\section{Poisson Ratio of the Piezoelectric Stack}
\label{app:Poisson}

\begin{figure}[!ht]  
\begin{center}  
\includegraphics[width=\columnwidth]{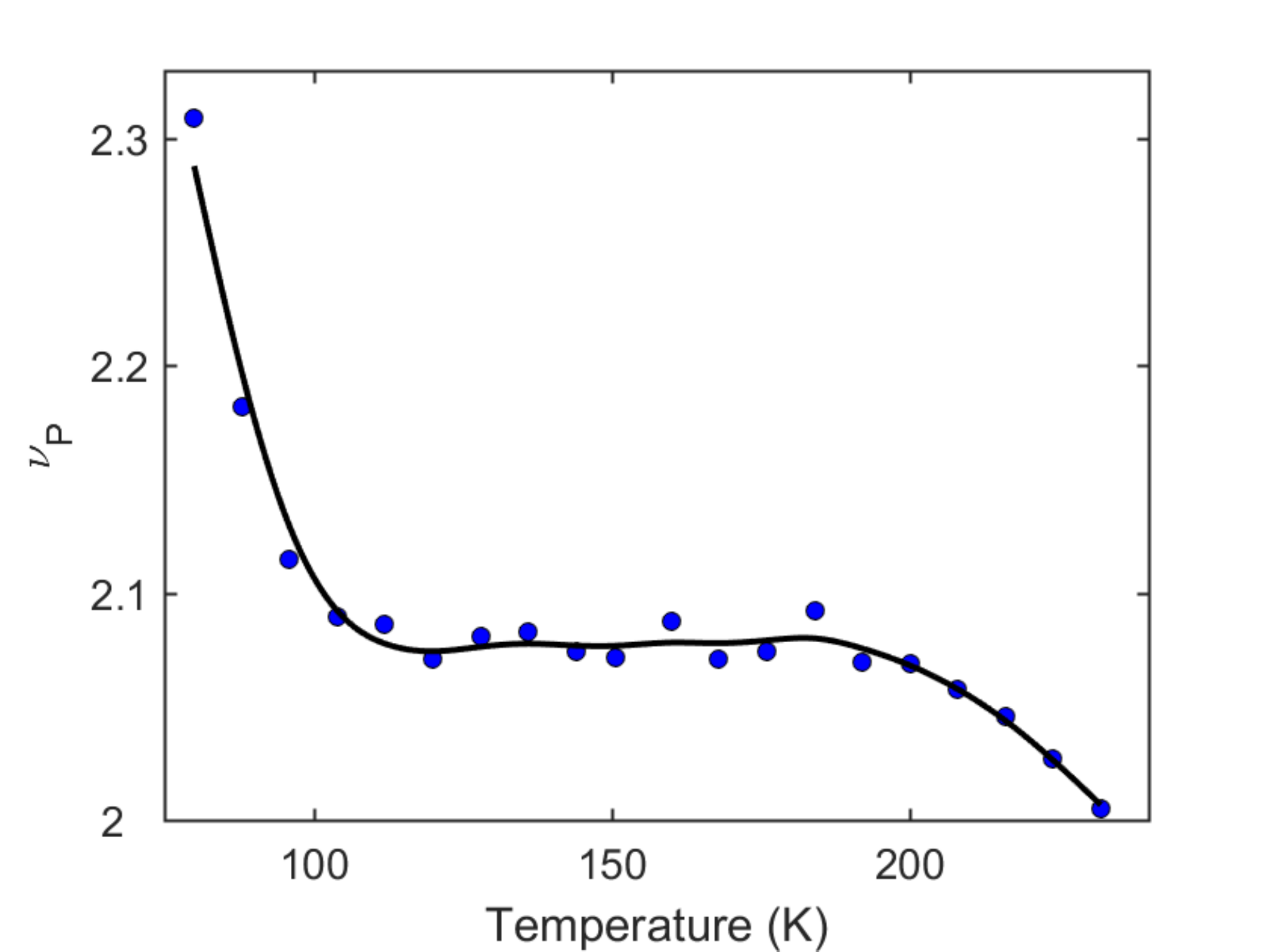}
\caption{\small \sl The Poisson ratio of the Piezomechanik ``PSt150/5x5/7 cryo 1''  piezoelectric stack as measured from two perpendicularly mounted strain gauges adhered to the surface of the piezoelectric with Devcon 5-Minute Epoxy.}
\label{fig:PZT_PoissonRatio}
\end{center}  
\end{figure}

Due to the large number of lock-ins required to measure each channel, it might not always be practical to measure both $\varepsilon_{x'x'}$ and $\varepsilon_{y'y'}$. 
However these two quantities are related by the Poisson ratio, $\nu_P$, of the piezoelectric stack ($\varepsilon_{y'y'}=-\nu_P\varepsilon_{x'x'}$). 
Characterization of the Poisson ratio therefore allows for the full symmetry decomposition of the measured strain from a single strain gauge measurement: for the specific example of the strain relevant to \cref{demo} of the main text, $\varepsilon_{xy}=\frac{1}{2}(\frac{\nu_P+1}{\nu_P})\varepsilon_{y'y'}$. 
In addition, typically $\nu_P>1$ so the measured signal along the $y'$ strain axis is larger and has a better signal to noise ratio. 
Therefore fully characterizing the Poisson ratio as a function of both temperature and frequency allows for a simpler experimental setup and cleaner signal. 
The values used in this experiment are provided in \cref{fig:PZT_PoissonRatio}.

\section{Heating from Piezoelectric Stacks}
\label{app:heat}
Driving the piezoelectric stacks at high frequencies and large amplitudes produces a large heat load near the sample.
Additionally, the piezoelectric stack has a very low thermal conductivity, which means that the heat generated is removed primarily via static exchange gas and contact leads.
The heat load for a given drive amplitude and frequency is temperature dependent as the capacitance of the piezoelectric stack (and thus the current required to drive it) is drastically reduced at low temperatures.
To account for this we use a measurement of the average resistivity of the sample being measured as an internal thermometer of the sample temperature (See \cref{app:Rxx}).
This is necessary for large drive frequencies and high temperatures.
For the 25~V peak-to-peak voltage used to drive the piezoelectric stack in this paper, significant heating onsets at frequencies above 2~kHz, which can be mitigated by reducing the amplitude at high frequencies at the cost of reducing the amplitude of the elastoresistivity signal.
A balance between these competing effects must be found for any given material. 
In our experience, it is best to focus on improving the quality of the electrical resistivity measurement of the sample, which in turn reduces the strains needed to measure the elastoresistivity and ultimately reduces the heating.

\section{Using Sample as a Resistive Thermometer}
\label{app:Rxx}
\begin{figure}[!ht]  
\begin{center}  
\includegraphics[width=\columnwidth]{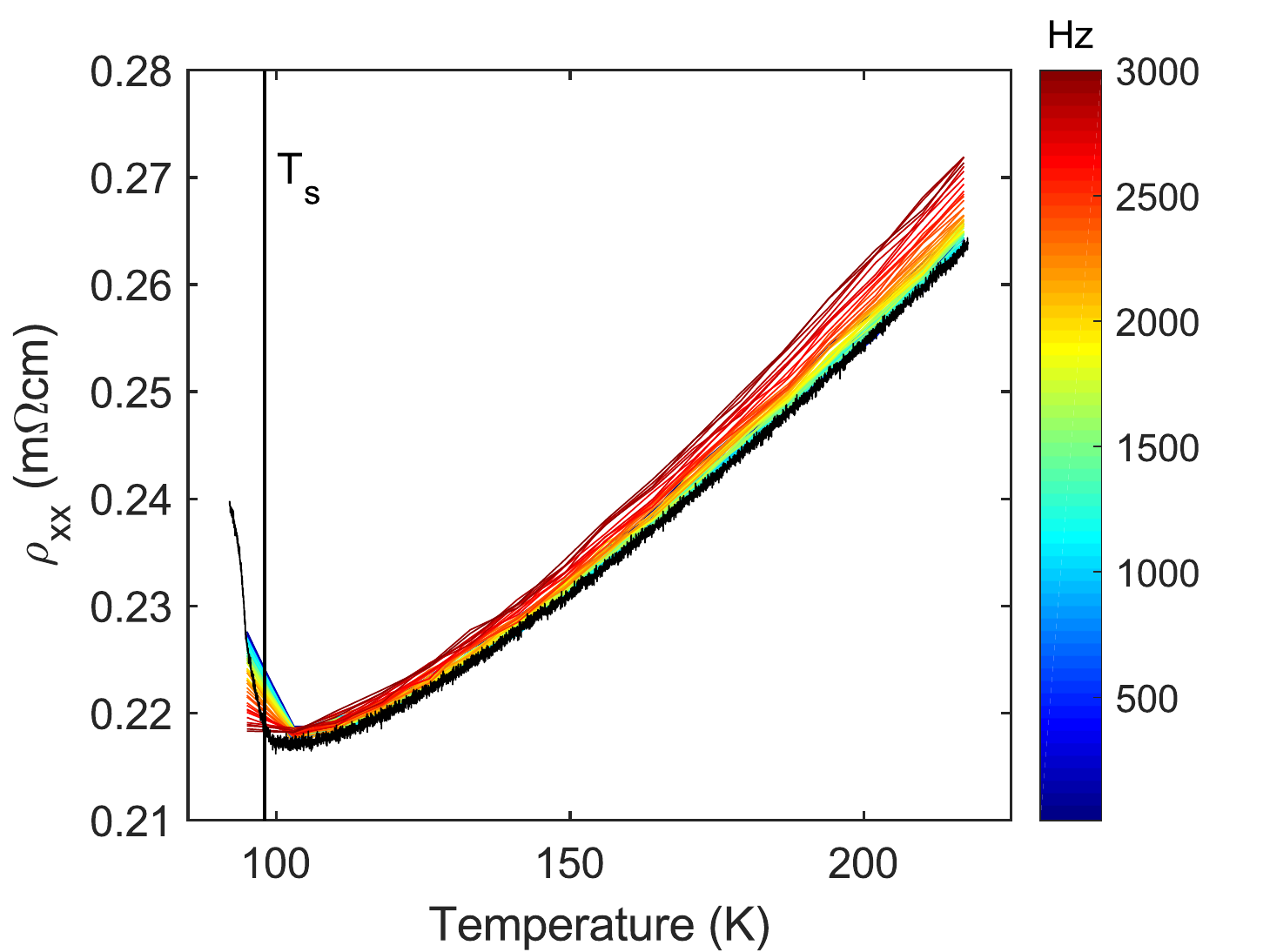}
\caption{\small \sl Longitudinal resistance of the sample as a function of strain frequency and cryostat heat exchanger temperature. As the strain frequency is increased, the piezoelectric and sample heat locally, causing the resistivity to increase from the value obtained with the piezo terminals shorted together (black curve).}
\label{fig:Rxx}  
\end{center}  
\end{figure}

\begin{figure}[!ht]  
\begin{center}  
\includegraphics[width=\columnwidth]{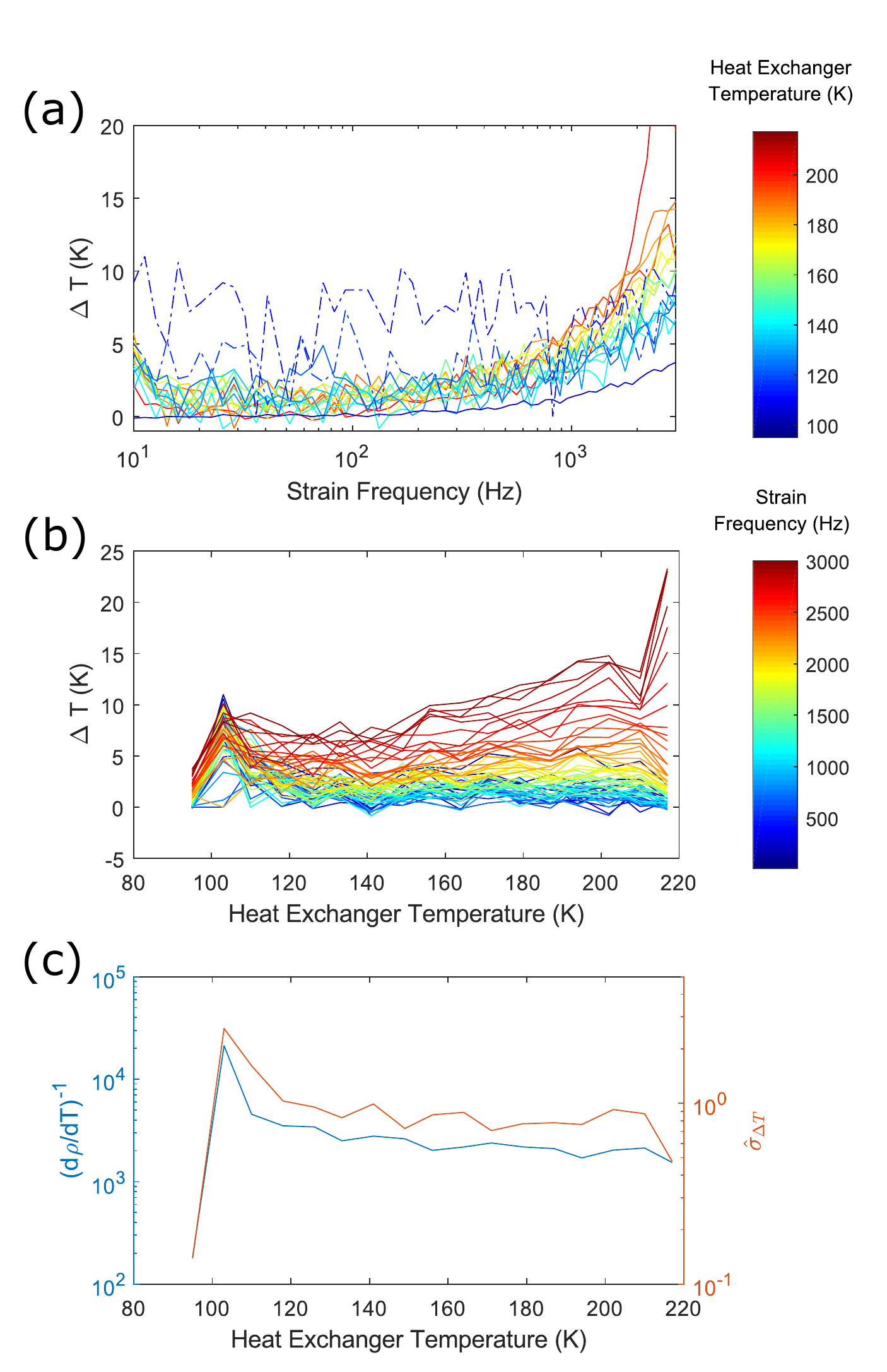}
\caption{
\small \sl Local heating at the piezo can be measured from the resistivity of the sample. 
The inferred temperature increase is defined as the temperature at which the unstrained sample (piezo terminals shorted) would exhibit the measured value of $\rho_{xx}$. 
Panel (a) shows the temperature increase with increasing strain frequency at fixed temperature (shown in color plot). 
The two sets of dashed lines correspond to temperatures of 103~K and 110~K, where the resistivity is significantly less temperature dependent. 
The same data is shown in panel (b) for varying temperature with frequency indicating the color. 
Panel (c) shows that the region where there is greatest variance of inferred temperature increase, $\hat{\sigma}_{\Delta T}$, (measured over strain frequencies less than 1~kHz to remove the effects of heating) also corresponds to the region where $d\rho/dT$ is closest to zero. 
It can be reasonably concluded heating estimates at these temperatures are not accurate, leading to the alternative estimate of the heating described in the text. 
}
\label{fig:heat}  
\end{center}  
\end{figure}

We use the longitudinal resistance of the sample as an internal thermometer.
In \cref{fig:Rxx} the black line corresponds to the resistivity of the sample when cooled in static exchange gas with the piezoelectric stack grounded.
Driving the piezoelectric stack at large frequencies and high temperatures causes the average resistivity values to increase at a given nominal exchange gas temperature.
For data above the structural transition, $T_s=98$~K, we map the resistivity changes to changes in sample temperature by comparing with the unheated trace.
Symmetry ensures there are no linear-order changes of this resistivity from the elastoresistivity response (the average of the applied sinusoidal strain is zero), and second order effects can be safely neglected in the limit of low strains.

When the resistivity becomes less temperature dependent, such as the case for our sample at temperatures close to 100~K, slight errors in the resistivity of the sample can lead to large estimates of heating.
As \cref{fig:Rxx} shows, the variability in our estimates of the heating is sharply peaked where the temperature derivative of the sample vanishes, between 100~K and 110~K.
Here, heating data from other temperatures can be an informative guide for how to estimate heating from the piezoelectric. 
Specifically, to estimate the heating at a given frequency of strain between 100~K and 110~K, we use an average of the heating from the piezoelectric at the same frequency from when the measurement was performed at 95~K and 118~K.

\section{Strain Transmission}
\label{app:strain}
\begin{figure}[!ht]  
\begin{center}  
\includegraphics[width=\columnwidth]{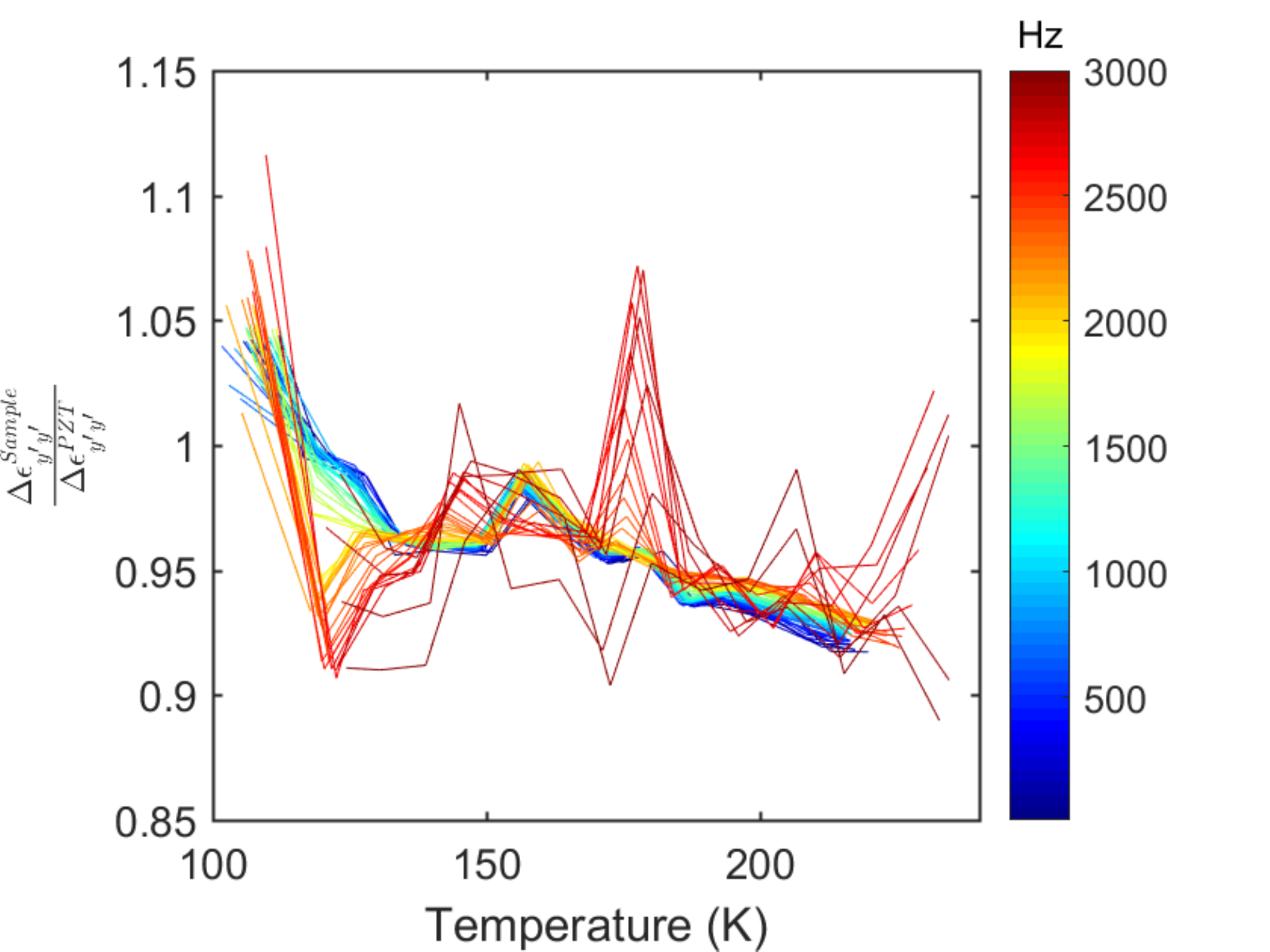}
\caption{\small \sl The ratio of strains measured by a strain gauge on the back surface of the piezoelectric and a second strain gauge mounted directly onto the sample, shown as a function of frequency and temperature. }
\label{fig:strain}  
\end{center}  
\end{figure}
An estimate of strain transmission through the sample is shown in \cref{fig:strain}.
The strain measured by the two strain gauges is typically within 5\% for most temperatures and frequencies and, the difference is never more than 12\%.
The strain transmission also only has a weak temperature and frequency dependence, justifying our approximation of using the strain gauge mounted on the piezoelectric as a proxy for the strain experienced by the sample. 

\section{Amplifier Details}
\label{appendix:amplifiers}

The current which charges the electrodes inside a piezoelectric stack scales with the frequency and amplitude of the voltage on the stack, and with the capacitance of the piezoelectric stack.
If sufficient voltages and currents cannot be provided to the piezoelectric stack, the quality of the elastoresistivity signal will deteriorate from either attenuation or harmonic distortion.
For this reason, selecting a suitable voltage source, typically a high voltage amplifier, is crucial for using techniques discussed in this paper.

\begin{figure}[!b]  
\begin{center}  
\includegraphics[width=\columnwidth]{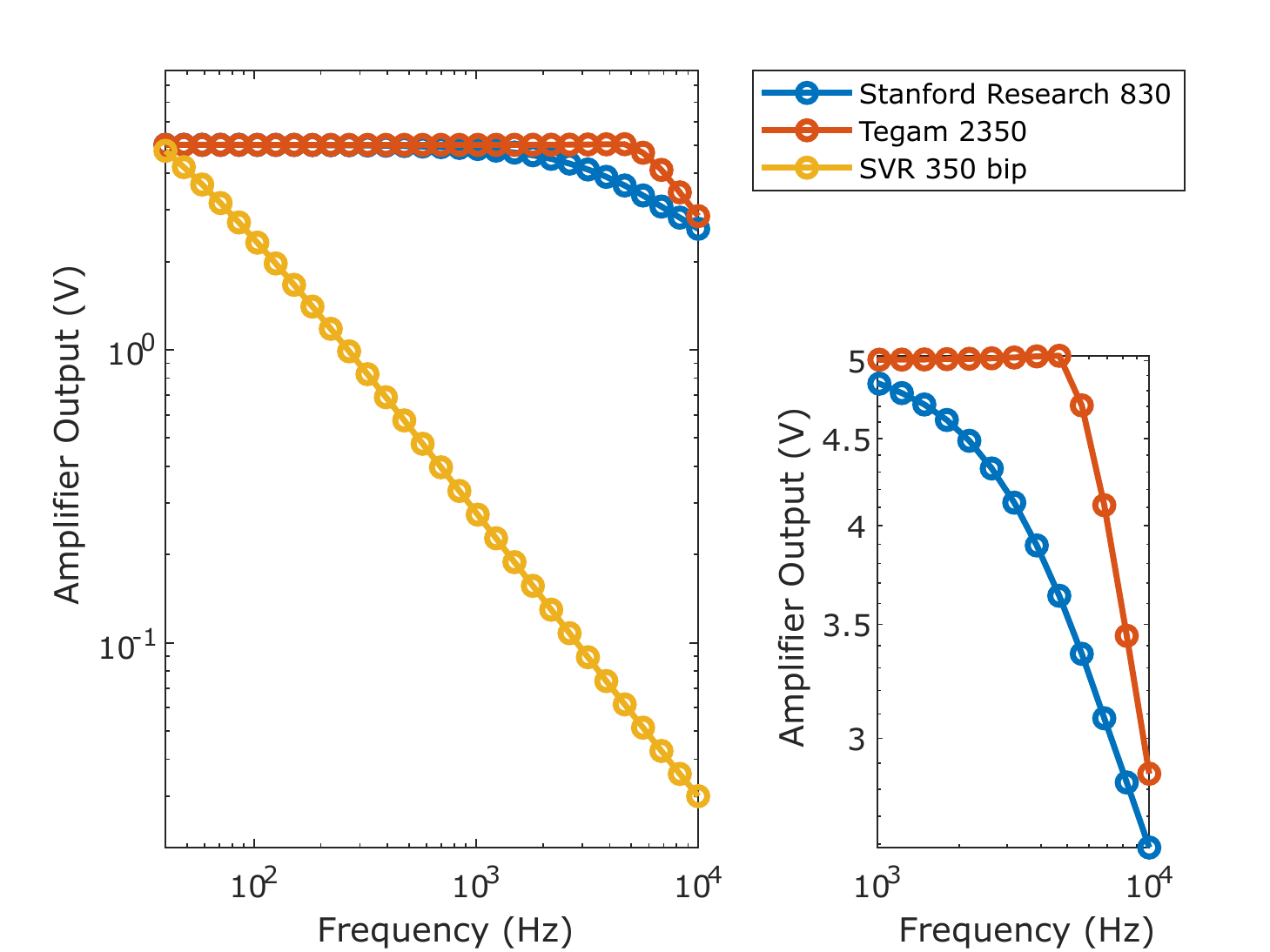}
\caption{\small \sl Performance three voltage sources when driving a Piezomechanik PSt ``PSt150/5x5/7 cryo 1'' stack. All three voltage sources are set to provide a 5~V rms sinusoidal voltage on the piezoelectric stack, and their output is measured by a lock-in amplifier through a 1:100 voltage divider. As described in the text, the Tegam amplifier approximates an ideal voltage source over a larger set of frequencies. }
\label{fig:Amplifier}  
\end{center}  
\end{figure}

\Cref{fig:Amplifier} shows the voltage output by three voltage sources amplifiers, an SVR-350-bip from Piezomechanik, the sine out of a Stanford Research 830 lock-in amplifier, and a 2350 amplifier from Tegam, when driving a piezoelectric stack at room temperature. 
Due to its large output current and frequency independent gain, the Tegam 2350 is able to drive a 5~V rms sine wave on the piezo to approximately 5~kHz.
Beyond this limit, the amplitude decreases and the raw trace on the oscilloscope shows increased distortion consistent saturation of the 40~mA current limit of the amplifier.
For comparison, the SVR-350-bip outputs a frequency dependent output at fixed input amplitude. 
As the output was observed on an oscilloscope to be always sinusoidal, it is inferred that either the gain or the output impedance vary with frequency.
At high frequencies, the higher output of the Tegam 2350 is clearly preferred.
At lower temperatures, the decreasing capacitance of the piezo makes it possible for this amplifier to drive even higher frequencies or amplitudes with fewer adverse effects from its current limitation.

When an amplifier is operating in the regime where it is current limited, the harmonic distortion of the output voltage necessitates lock-in methods to measure the strain gauge and sample response to only the strain at a particular frequency.

\section{Correcting for Lock-In Amplifier Transfer Function and Current Source}
\label{app:TransferFunction}
The lock-in amplifier provides gain and a low pass filter to any amplitude modulation signal arising from the elastoresistivity of the sample. 
The four-pole filter for the digital signal processing lock-in amplifier used here is specified to take the form
\begin{equation}
T_{\tau,\text{24 dB/oct}}(\omega)=\left(\frac{1 -  i 2\pi\omega\tau }{1 + 4\pi^2\omega^2\tau^2}\right)^4,
\label{eq:TFunc}
\end{equation}
where $\tau$ is the time constant specified at the front panel of the instrument.

\begin{figure}[!b]  
\begin{center}  
\includegraphics[width=\columnwidth]{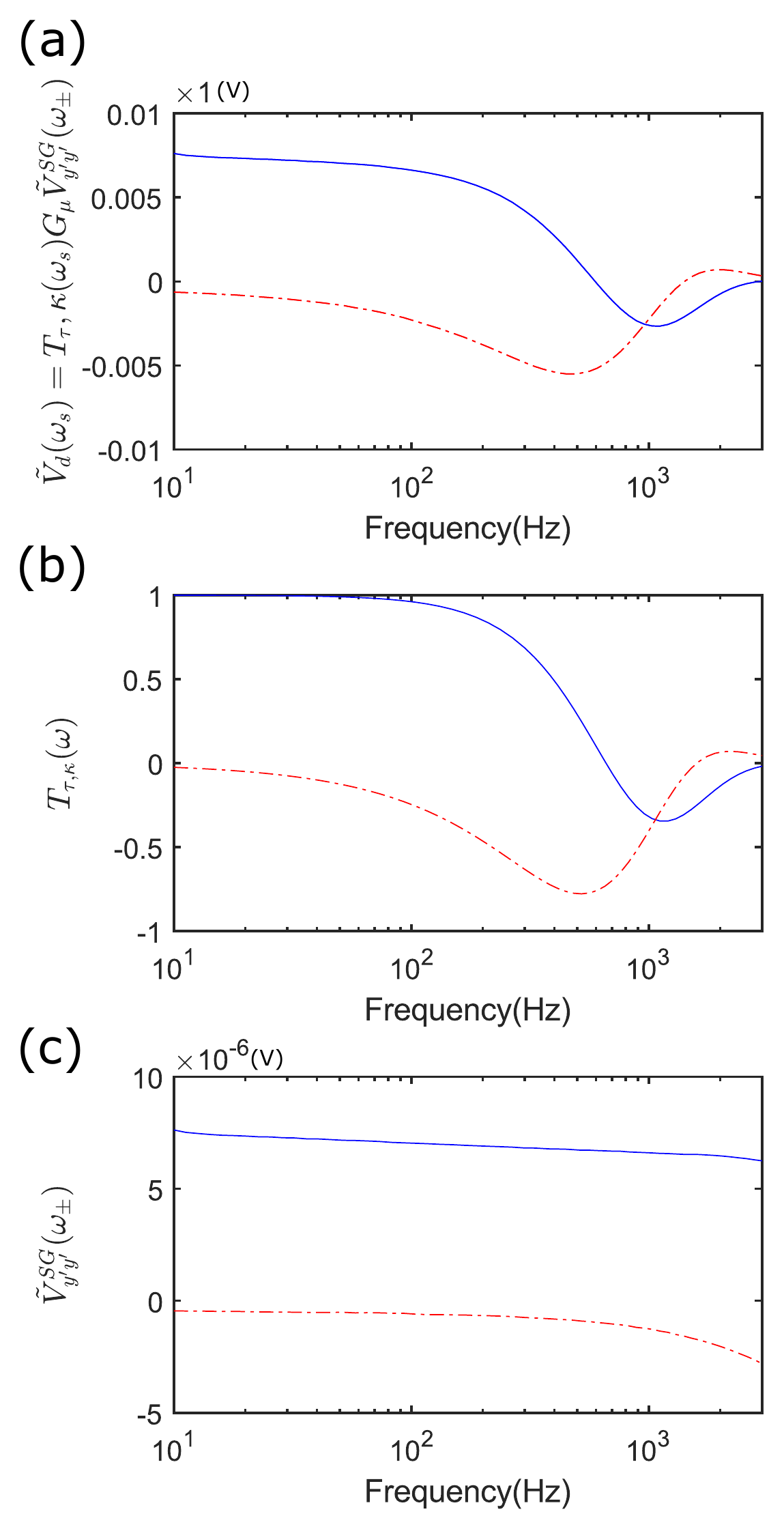}
\caption{Response of a strain gauge mounted along the poling direction of the piezoelectric stack, at 133~K, as a function of strain frequency. Strain is induced through a 25~V peak-to-peak amplitude sine wave to the piezoelectric stack while a current of 1~mA rms flows through a Wheatstone bridge consisting of the strain gauge and 3 other resistors. 
(a) The real (blue) and imaginary (red) components of the $\omega_s$ component of the demodulated voltage from the strain gauge, as output by a lock-in connected to the differential terminals of the Wheatstone bridge.
(b) Transfer function of the 4-pole digital filter calculated from \cref{eq:TFunc} for a time constant of 100~\micro\second.
(c) Inverting the gain from the lock-in amplifier ($G_\mu=1000$) and the effects of the transfer function yields the total sideband amplitude voltage.}
\label{fig:TransferFunction}  
\end{center}  
\end{figure}

As outlined in \cref{sec:electronics}, obtaining the correct values of the elastoresistivity sidebands requires inverting the effects of this filter. 
A representative example of this process is shown in \cref{fig:TransferFunction}, which shows the response of a strain gauge measured at 133~K using the dual lock-in method described in the text.
The $\omega_s$ component of the demodulated signal $V_d$ appears to have frequency dependence in both amplitude and phase, but this is shown to arise predominantly from the transfer function of the first processing step, in which a lock-in amplifier demodulates and filters this signal.
Indeed, once the effects of the transfer function are inverted, it is clear that the amplitude and phase are stable for strain frequencies to 3~kHz; the slight increase in the imaginary component might be due to either self heating or capacitive coupling, but does not meaningfully affect the measurement. 
In practice, when the time constants are set identically for the first demodulation and filtering step in each measurement channel, these transfer functions can be neglected as they cancel out when appropriate ratios are taken of the changes in resistivity between sample and strain gauge.

Another important consideration comes from the behavior of the current source.
We observed a frequency dependent shift between the supplied current and the reference frequency signal from our current source, creating an artificial phase shift of  approximately 10 degrees that does not come from the cryostat wiring or sample. 
In such a configuration, the voltage from the sample is partially rotated in phase space and the modulation of the in-quadrature amplitude is not passed on to the subsequent instruments. 
This is accounted for by a small correction to \cref{eq:deltarho}: when the quantity $\rnorm$ is simultaneously measured through identical wiring and an identical current source, the in-phase amplitude of the corresponding voltage is used in place of the total amplitude.

\section{Alternative techniques for measuring dynamical elastoresistivity}
\label{app:ames}

During the course of preparing this manuscript we became aware of another experiment that also measures the frequency dependence of the elastoresistivity response.\cite{Gil_Dynamic_2017}
Rather than applying a modulation-demodulation technique, a DC current was used to excite a sample experiencing an AC strain.
In this alternative measurement, the elastoresistivity of the sample was obtained from a lock-in measurement at the same frequency as the strain.
While this alternative technique can in principle measure the dynamic elastoresistivity, in practice certain experimental realities can limit the effectiveness. In particular, using a DC current allows a capacitive coupling between the current driving the piezoelectric and the elastoresistivity signal as both occur at the same frequency. 
This effect must be accounted for with careful subtraction of the capacitive coupling background, which can be difficult for less resistive samples and high frequencies of strain.
By using the demodulation technique presented here, capacitive crosstalk from the piezoelectric enters measured signals at $\omega_s$, distinct from our elastoresistivity signal of interest which lies at $\omega_c\pm\omega_s$.
Furthermore, some DC current sources are not generally suited to driving loads that vary at high frequency; we have found that these current sources can artificially attenuate the measured signal at high strain frequencies, further motivating use of the demodulation technique that we describe in this paper, which necessitates an AC current source.

\end{appendices}

\bibstyle{aps}
\bibliography{References}

\end{document}